\documentclass[%
 reprint,
superscriptaddress,
nofootinbib,
 amsmath,amssymb,
 aps,
 prd,
 showkeys,
floatfix,
]{revtex4-2}

\usepackage{graphicx}% Include figure files
\usepackage{dcolumn}% Align table columns on decimal point
\usepackage{bm}% bold math
\usepackage{amssymb}   % for math
\usepackage{aas_macros}
\usepackage{hyperref}% add hypertext capabilities
\usepackage{amsmath}
\usepackage{accents}
\usepackage[caption=false]{subfig}
\newcommand{\HI}{H\,{\sc i}}

\newcommand*\conj[1]{\overline{#1}}

\newcommand\fdg{\hbox{$.\!\!^\circ$}}

\newcommand\arcmin{\hbox{$^\prime$}}

\newcommand\arcdeg{\mbox{$^\circ$}}
\graphicspath{{./}{figures/}}

\begin{document}

\preprint{APS/123-QED}

\title{Detection of Cosmic Structures using the Bispectrum Phase. II. \\ First Results from Application to Cosmic Reionization Using the Hydrogen Epoch of Reionization Array}

%% you may use \\ to force a line break if

% Corresponding author
\author{Nithyanandan Thyagarajan}
\thanks{Nithyanandan Thyagarajan is a Jansky Fellow of the National Radio Astronomy Observatory}
\email{\\ t\_nithyanandan@nrao.edu, nithyanandan.t@gmail.com}
\homepage{\\ https://tnithyanandan.wordpress.com/}
\affiliation{National Radio Astronomy Observatory, 1003 Lopezville Road, Socorro, NM 87801, USA}
\affiliation{School of Earth and Space Exploration, Arizona State University, Tempe, AZ, USA}

% Contributing authors
\author{Chris L. Carilli}
\affiliation{National Radio Astronomy Observatory, 1003 Lopezville Road, Socorro, NM 87801, USA}
\affiliation{Astrophysics Group, Cavendish Laboratory, University of Cambridge, Cambridge CB3 0HE, UK}

\author{Bojan Nikolic}
\affiliation{Astrophysics Group, Cavendish Laboratory, University of Cambridge, Cambridge CB3 0HE, UK}

\author{James Kent}
\affiliation{Astrophysics Group, Cavendish Laboratory, University of Cambridge, Cambridge CB3 0HE, UK}

\author{Andrei  Mesinger}
\affiliation{Scuola Normale Superiore, 56126 Pisa, PI, Italy}

\author{Nicholas S. Kern}
\affiliation{Department of Astronomy, University of California, Berkeley, CA}

\author{Gianni  Bernardi}
\affiliation{Department of Physics and Electronics, Rhodes University, PO Box 94, Grahamstown, 6140, South Africa}
\affiliation{INAF-Istituto di Radioastronomia, via Gobetti 101, 40129 Bologna, Italy}
\affiliation{The South African Radio Astronomy Observatory (SARAO), 2 Fir Street, Black River Park, Observatory, 7925, South Africa}

\author{Siyanda Matika}
\affiliation{Department of Physics and Electronics, Rhodes University, PO Box 94, Grahamstown, 6140, South Africa}

% Builders List

\author{Zara  Abdurashidova}
\affiliation{Department of Astronomy, University of California, Berkeley, CA}

\author{James E. Aguirre}
\affiliation{Department of Physics and Astronomy, University of Pennsylvania, Philadelphia, PA}

\author{Paul  Alexander}
\affiliation{Astrophysics Group, Cavendish Laboratory, University of Cambridge, Cambridge CB3 0HE, UK}

\author{Zaki S. Ali}
\affiliation{Department of Astronomy, University of California, Berkeley, CA}

\author{Yanga  Balfour}
\affiliation{The South African Radio Astronomy Observatory (SARAO), 2 Fir Street, Black River Park, Observatory, 7925, South Africa}

\author{Adam P. Beardsley}
\affiliation{School of Earth and Space Exploration, Arizona State University, Tempe, AZ}
\affiliation{NSF Astronomy \& Astrophysics Postdoctoral Fellow}

\author{Tashalee S. Billings}
\affiliation{Department of Physics and Astronomy, University of Pennsylvania, Philadelphia, PA}

\author{Judd D. Bowman}
\affiliation{School of Earth and Space Exploration, Arizona State University, Tempe, AZ}

\author{Richard F. Bradley}
\affiliation{National Radio Astronomy Observatory, Charlottesville, VA}

\author{Jacob  Burba}
\affiliation{Department of Physics, Brown University, Providence, RI}

\author{Steve  Carey}
\affiliation{Cavendish Astrophysics, University of Cambridge, Cambridge, UK}

\author{Carina  Cheng}
\affiliation{Department of Astronomy, University of California, Berkeley, CA}

\author{David R. DeBoer}
\affiliation{Department of Astronomy, University of California, Berkeley, CA}

\author{Matt  Dexter}
\affiliation{Department of Astronomy, University of California, Berkeley, CA}

\author{Eloy  de~Lera~Acedo}
\affiliation{Cavendish Astrophysics, University of Cambridge, Cambridge, UK}

\author{Joshua S. Dillon}
\affiliation{Department of Astronomy, University of California, Berkeley, CA}
\affiliation{NSF Astronomy \& Astrophysics Postdoctoral Fellow}

\author{John  Ely}
\affiliation{Cavendish Astrophysics, University of Cambridge, Cambridge, UK}

\author{Aaron  Ewall-Wice}
\affiliation{Department of Physics, Massachusetts Institute of Technology, Cambridge, MA}

\author{Nicolas  Fagnoni}
\affiliation{Cavendish Astrophysics, University of Cambridge, Cambridge, UK}

\author{Randall  Fritz}
\affiliation{The South African Radio Astronomy Observatory (SARAO), 2 Fir Street, Black River Park, Observatory, 7925, South Africa}

\author{Steven R. Furlanetto}
\affiliation{Department of Physics and Astronomy, University of California, Los Angeles, CA}

\author{Kingsley  Gale-Sides}
\affiliation{Cavendish Astrophysics, University of Cambridge, Cambridge, UK}

\author{Brian  Glendenning}
\affiliation{National Radio Astronomy Observatory, 1003 Lopezville Road, Socorro, NM 87801, USA}

\author{Deepthi  Gorthi}
\affiliation{Department of Astronomy, University of California, Berkeley, CA}

\author{Bradley  Greig}
\affiliation{School of Physics, University of Melbourne, Parkville, VIC 3010, Australia}

\author{Jasper  Grobbelaar}
\affiliation{The South African Radio Astronomy Observatory (SARAO), 2 Fir Street, Black River Park, Observatory, 7925, South Africa}

\author{Ziyaad  Halday}
\affiliation{The South African Radio Astronomy Observatory (SARAO), 2 Fir Street, Black River Park, Observatory, 7925, South Africa}

\author{Bryna J. Hazelton}
\affiliation{Department of Physics, University of Washington, Seattle, WA}
\affiliation{eScience Institute, University of Washington, Seattle, WA}

\author{Jacqueline N. Hewitt}
\affiliation{Department of Physics, Massachusetts Institute of Technology, Cambridge, MA}

\author{Jack  Hickish}
\affiliation{Department of Astronomy, University of California, Berkeley, CA}

\author{Daniel C. Jacobs}
\affiliation{School of Earth and Space Exploration, Arizona State University, Tempe, AZ}

\author{Austin  Julius}
\affiliation{The South African Radio Astronomy Observatory (SARAO), 2 Fir Street, Black River Park, Observatory, 7925, South Africa}

\author{Joshua  Kerrigan}
\affiliation{Department of Physics, Brown University, Providence, RI}

\author{Piyanat  Kittiwisit}
\affiliation{School of Chemistry and Physics, University of KwaZulu-Natal, Westville Campus, Durban, South Africa}

\author{Saul A. Kohn}
\affiliation{Department of Physics and Astronomy, University of Pennsylvania, Philadelphia, PA}

\author{Matthew  Kolopanis}
\affiliation{School of Earth and Space Exploration, Arizona State University, Tempe, AZ}

\author{Adam  Lanman}
\affiliation{Department of Physics, Brown University, Providence, RI}

\author{Paul  La~Plante}
\affiliation{Department of Physics and Astronomy, University of Pennsylvania, Philadelphia, PA}

\author{Telalo  Lekalake}
\affiliation{The South African Radio Astronomy Observatory (SARAO), 2 Fir Street, Black River Park, Observatory, 7925, South Africa}

\author{David  Lewis}
\affiliation{School of Earth and Space Exploration, Arizona State University, Tempe, AZ}

\author{Adrian  Liu}
\affiliation{Department of Physics and McGill Space Institute, McGill University, 3600 University Street, Montreal, QC H3A 2T8, Canada}

\author{David  MacMahon}
\affiliation{Department of Astronomy, University of California, Berkeley, CA}

\author{Lourence  Malan}
\affiliation{The South African Radio Astronomy Observatory (SARAO), 2 Fir Street, Black River Park, Observatory, 7925, South Africa}

\author{Cresshim  Malgas}
\affiliation{The South African Radio Astronomy Observatory (SARAO), 2 Fir Street, Black River Park, Observatory, 7925, South Africa}

\author{Matthys  Maree}
\affiliation{The South African Radio Astronomy Observatory (SARAO), 2 Fir Street, Black River Park, Observatory, 7925, South Africa}

\author{Zachary E. Martinot}
\affiliation{Department of Physics and Astronomy, University of Pennsylvania, Philadelphia, PA}

\author{Eunice  Matsetela}
\affiliation{The South African Radio Astronomy Observatory (SARAO), 2 Fir Street, Black River Park, Observatory, 7925, South Africa}

\author{Mathakane  Molewa}
\affiliation{The South African Radio Astronomy Observatory (SARAO), 2 Fir Street, Black River Park, Observatory, 7925, South Africa}

\author{Miguel F. Morales}
\affiliation{Department of Physics, University of Washington, Seattle, WA}

\author{Tshegofalang  Mosiane}
\affiliation{The South African Radio Astronomy Observatory (SARAO), 2 Fir Street, Black River Park, Observatory, 7925, South Africa}

\author{Abraham R. Neben}
\affiliation{Department of Physics, Massachusetts Institute of Technology, Cambridge, MA}

\author{Aaron R. Parsons}
\affiliation{Department of Astronomy, University of California, Berkeley, CA}

\author{Nipanjana  Patra}
\affiliation{Department of Astronomy, University of California, Berkeley, CA}

\author{Samantha  Pieterse}
\affiliation{The South African Radio Astronomy Observatory (SARAO), 2 Fir Street, Black River Park, Observatory, 7925, South Africa}

\author{Jonathan C. Pober}
\affiliation{Department of Physics, Brown University, Providence, RI}

\author{Nima  Razavi-Ghods}
\affiliation{Cavendish Astrophysics, University of Cambridge, Cambridge, UK}

\author{Jon  Ringuette}
\affiliation{Department of Physics, University of Washington, Seattle, WA}

\author{James  Robnett}
\affiliation{National Radio Astronomy Observatory, 1003 Lopezville Road, Socorro, NM 87801, USA}

\author{Kathryn  Rosie}
\affiliation{The South African Radio Astronomy Observatory (SARAO), 2 Fir Street, Black River Park, Observatory, 7925, South Africa}

\author{Peter  Sims}
\affiliation{Department of Physics, Brown University, Providence, RI}

\author{Craig  Smith}
\affiliation{The South African Radio Astronomy Observatory (SARAO), 2 Fir Street, Black River Park, Observatory, 7925, South Africa}

\author{Angelo  Syce}
\affiliation{The South African Radio Astronomy Observatory (SARAO), 2 Fir Street, Black River Park, Observatory, 7925, South Africa}

\author{Peter K.~G. Williams}
\affiliation{Center for Astrophysics | Harvard \& Smithsonian, Cambridge, MA}
\affiliation{American Astronomical Society, Washington, DC}

\author{Haoxuan  Zheng}
\affiliation{Department of Physics, Massachusetts Institute of Technology, Cambridge, MA}

\date{\today}% It is always \today, today,

\begin{abstract}

Characterizing the epoch of reionization (EoR) at $z\gtrsim 6$ via the redshifted 21~cm line of neutral Hydrogen (\HI) is critical to modern astrophysics and cosmology, and thus a key science goal of many current and planned low-frequency radio telescopes. The primary challenge to detecting this signal is the overwhelmingly bright foreground emission at these frequencies, placing stringent requirements on the knowledge of the instruments and inaccuracies in analyses. Results from these experiments have largely been limited not by thermal sensitivity but by systematics, particularly caused by the inability to calibrate the instrument to high accuracy. The interferometric bispectrum phase is immune to antenna-based calibration and errors therein, and presents an independent alternative to detect the EoR \HI\ fluctuations while largely avoiding calibration systematics. Here, we provide a demonstration of this technique on a subset of data from the Hydrogen Epoch of Reionization Array (HERA) to place approximate constraints on the brightness temperature of the intergalactic medium (IGM). From this limited data, at $z=7.7$ we infer ``$1\sigma$'' upper limits on the IGM brightness temperature to be $\le 316$~``pseudo''~mK at $\kappa_\parallel=0.33\,\textrm{``pseudo''}\,h$~Mpc$^{-1}$ (data-limited) and $\le 1000$~``pseudo''~mK at $\kappa_\parallel=0.875\,\textrm{``pseudo''}\,h$~Mpc$^{-1}$ (noise-limited). The ``pseudo'' units denote only an approximate and not an exact correspondence to the actual distance scales and brightness temperatures. By propagating models in parallel to the data analysis, we confirm that the dynamic range required to separate the cosmic \HI\ signal from the foregrounds is similar to that in standard approaches, and the power spectrum of the bispectrum phase is still data-limited (at $\gtrsim 10^6$ dynamic range) indicating scope for further improvement in sensitivity as the array build-out continues.

\end{abstract}

%Use showkeys class option if keyword display desired

\maketitle

% \tableofcontents

\section{Introduction} \label{sec:intro}

The epoch of reionization (EoR) is an important period in the evolution of the Universe, which is characterized by nonlinear growth of matter density perturbations and astrophysical evolution. It marks the point at which structure formation directly affected every baryon in the intergalactic medium (IGM). Large areas of the IGM were photo-ionized by the radiation produced in galaxies and it eventually ended when the IGM had become completely ionized ($z\sim 6$). Thus, the Universe transitioned from being fully neutral to being fully ionized. 

The 21~cm spin-flip transition of neutral hydrogen (\HI) during the EoR can provide insight into the formation, growth and evolution of structure in the Universe, the nature of the first stars and galaxies, and their impact on the physics of the IGM \cite[e.g.][]{bar07,loe13,zar13,fan06}. Using the redshifted 21~cm line of \HI\ has the following advantages \cite{fur06}: (a) Being a cosmological spectral line, the redshift information can be used to trace the full three-dimensional ionization history, (b) Being the most abundant element in the IGM, it directly probes the IGM which constitutes a major fraction of baryonic matter, and (c) Being a forbidden transition, it does not saturate easily and is thus sensitive to various stages of reionization. Therefore, it is considered as one of the most promising and direct probes of the EoR \cite[see e.g.][]{gne97,mad97,sha99,toz00,gne04,fur06,mor10,pri12}.

To study the IGM structures using redshifted 21~cm line from \HI\ during the EoR, numerous interferometer-based experiments at low radio frequencies have and will become operational. These include the Murchison Widefield Array \cite[MWA;][]{lon09,bow13,tin13,bea19}, the Donald~C.~Backer Precision Array for Probing the Epoch of Reionization \cite[PAPER;][]{par10}, the Low Frequency Array \cite[LOFAR;][]{van13}, the Giant Metrewave Radio Telescope EoR experiment \cite[GMRT;][]{pac13}, the Hydrogen Epoch of Reionization Array \cite[HERA\footnote{\url{http://reionization.org/}};][]{deb17}, and the Square Kilometre Array \cite[SKA\footnote{\url{https://www.skatelescope.org/}};][]{mel13}. Many of these instruments, especially those belonging to the current generation, have sensitivity sufficient only for a statistical detection of the EoR signal by estimating the spatial power spectrum of the redshifted \HI\ spin temperature fluctuations \cite{bea13,thy13} and not for a three-dimensional tomographic imaging of the ionization structures yet. 

Almost all these low frequency experiments have to contend with the challenge of achieving a daunting spectral dynamic range, typically $\gtrsim 10^5:1$, posed by the extremely bright sources of radio emission from foreground objects including the Galaxy and the extragalactic radio sources which are many orders of magnitude brighter than the cosmological \HI\ signal. However, these foregrounds typically have smooth spectra as they arise predominantly from synchrotron and free-free emission mechanisms that result in a smooth continuum. In contrast, the imprinted \HI\ signatures may be faint but are expected to exhibit sharp fluctuating signatures in their spectrum. Therefore, distinguishing this spectral contrast forms the primary basis for separating the \HI\ signal from the foregrounds in most of the approaches that rely either on tomography or power spectrum \cite{dat09,dat10,liu10,par12b,ved12,pac13,zhe14,thy15a,thy15b,thy16,tro16}.  

Avoiding spectral leakage from these bright foregrounds is critical for the success of these experiments. It not only requires that the instruments be designed and characterized to a fractional accuracy better than $10^{-5}$ \cite[see e.g.][]{thy16,neb16,ewa16,deb17,patra18,fag19}, but also requires analysis methods, particularly involving calibration, to be equally accurate or better \cite[see e.g.][]{dat10,barry16,tro16,patil17}. It is now evident that the results from a majority of these experiments and associated analyses are predominantly limited by contamination from systematics, rather than by thermal noise \cite[see e.g.][]{pac13,dil15,bea16,patil17,che18,barry19b,kol19}. Therefore, analyzing more data is not helpful in improving sensitivity unless substantial improvements are made in the instrument and in the analyses to mitigate these systematics. Numerous sophisticated schemes are indeed being developed to address the calibration challenge \cite[see e.g.][]{ewa17,sie17,dil18,oro19}.

While a tentative claimed detection of deep absorption in the redshift window spanning \textit{the Dark Ages} \cite{bow18} has raised interest in detecting \HI\ 21cm emission from these very early cosmic epochs, the challenges faced by all classes of redshifted 21~cm experiments  have firmly established that any detection of cosmic \HI\ requires independent and credible confirmation. Besides independent instruments, we are witnessing the development of a number of novel techniques that probe the EoR from various angles \cite[see e.g.][]{wat14,wat15,wat19,maj18,kit18a,kit18b,patwa19,tro19a,tro19b,gor19}. A vast majority of these techniques, however, are still dependent on high accuracy in instrument calibration, the errors in which are currently the primary limitation to a successful EoR detection.

In interferometry, the bispectrum phase (also known as closure phase in radio interferometry) is immune to errors in direction-independent, antenna-based calibration as they are independent of the complex antenna gains \cite{jen58}. Therefore, it has found successful applications in many fields where interferometric calibration is challenging, such as deciphering complex structures on stellar surfaces and their surroundings \cite[see e.g.][]{mon07a,mon07b}, and more recently in the imaging of the shadow of the super-massive black hole at the center of M87 \cite{eht19-1,eht19-2,eht19-3,eht19-4,eht19-5,eht19-6}. 

This paper is one in a series of related papers, the others being \cite{thy18,car18,car20,thy20a}. Recently, a new approach to statistically detect the \HI\ spin temperature fluctuations from the EoR employing the concept of bispectrum phase was presented \cite{thy18}, which promises to sidestep to a large extent the calibration challenge and detect EoR \HI\ fluctuations in the presence of strong foreground emission. The utility of bispectrum phase in diagnosing radio interferometer arrays was illustrated in \cite{car18}. In a companion paper \cite[hereafter Paper I]{thy20a}, the mathematical foundations of the bispectrum phase for detecting faint spectral line fluctuations from cosmic structures is presented. In parallel work \cite{car20}, a subset of data (same as in this paper) from the HERA observations and the parallel modeling effort is presented. However, in order to be self-contained, all the relevant information about the data and the modeling effort pertinent to this work is summarized in this paper as well. This paper is the fifth in this series. Here, we provide a first demonstration of the technique as well as first results from data obtained using the HERA telescope using the basis in \cite{thy18,car18,car20,thy20a}.

This paper is organized as follows. In Section~\ref{sec:BSP-context}, we state the scope and context of our analysis and interpretation. Section~\ref{sec:Data} describes the data used in this analysis, Section~\ref{sec:Modeling} elaborates the forward-modeling we have employed to verify and interpret our results. In Section~\ref{sec:Analysis}, we describe in detail the analysis steps and present the results in Section~\ref{sec:Results}. We summarize this work and discuss the prospects of a larger analysis effort in Section~\ref{sec:Summary}. 

\section{Bispectrum Phase in the context of detecting \HI\ Fluctuations}\label{sec:BSP-context}

The authors of \cite{thy18} have shown the potential for using the bispectrum phase to detect the EoR signal using a power spectrum methodology very similar to the delay spectrum approach \cite{par12b}. Numerous works \cite[][and references therein]{car18} have reiterated the reasons the bispectrum phase is impervious to direction-independent antenna-based gains, and therefore, also to the corresponding calibration and errors therein \cite{jen58}. They all reinforce the utility of the approach to eliminate potentially a major source of systematic artefact that is believed to be affecting current EoR power spectrum limits. 

In Paper I, the information contained in the bispectrum phase and its power spectrum, as well as its relation to the standard power spectrum approach is investigated in detail. Under the assumption that the cosmological signal strength is small compared to the foreground contributions that dominate the bispectrum phase, the fluctuations in the bispectrum phase can be treated up to linear-order terms and they have approximate correspondence with the power spectrum of the cosmological signal. In low frequency EoR experiments, this assumption and the approximation to linear-order terms are justified. 

Nevertheless, the aim of this paper is not to identify or interpret the results in strict cosmological terms. Such interpretation will require extensive forward-modeling over a wide range of parameter space. Ideally, the bispectrum phase is a useful interferometric quantity very  sensitive to structures on the sky providing a robust avenue to discern the presence of EoR \HI\ spin temperature fluctuations against the foregrounds. And here, we only present our analysis using the bispectrum phase as an independent step towards detecting the EoR \HI\ signal against the null hypothesis that such a signal is absent.

For this work, we use cosmological parameters from \cite{planck15xiii} with $H_0=100\,h$~km~s$^{-1}$~Mpc$^{-1}$. 

\section{Data}\label{sec:Data}

We use a subset of the Phase I commissioning data of the HERA project (denoted as H1C-IDR2.0) obtained in 2018 with 61 dishes. The layout of this array can be seen in Figure 2 in \cite{car18}. Of these, data from only 50 unflagged antennas were subsequently selected. The data used here spans 22 minutes repeated over 18 nights on each of the two fields, one centered on the transit of Fornax A and the other starting at RA=01h36m, which will be hereafter referred to as the Fornax and J0136-30 fields respectively. The data consists of 1024 spectral channels each 97656.25~Hz wide for a total bandwidth of 100~MHz and a temporal resolution of 10.7~s. The preliminary data analysis including calibration and imaging processes are described in \cite{car20,ker20b}, which are subsequently used to compare the data and models to each other. The data and the models are in good agreement with each other indicating that the HERA instrument is performing reasonably as per expectations and that our models capture most of the important features seen in the data \cite{car20}. However, this work treads a different path and essentially uses the raw pre-calibration data and therefore, requires only minimal pre-processing prior to this analysis. In this paper, we restrict our analysis to equilateral antenna triads of separation 29.2~m.

\section{Modeling}\label{sec:Modeling}

For a quantitative interpretation the bipsectrum phase results, we have carried out detailed modeling to create forward-models of two specific realizations of the foregrounds (one for each field), and one realization each of the noise and the fiducial EoR \HI\ signal to understand the characteristics seen in the data. We have used the Precision Radio Interferometry Simulator \cite[PRISim\footnote{PRISim is publicly available for use under the MIT license at \url{https://github.com/nithyanandan/PRISim}};][]{PRISim_software} to model the visibilities based on the observing and instrument configurations of the HERA telescope. The simulations consist of the key components, including noise, summarized further below in this section. Sky modeling, in the context of our bispectrum phase delay spectrum analysis, is required only for forward-modeling, i.e., as a comparison to the measurements. The bispectrum phase technique does not require direction-independent antenna-based calibration, and hence the required accuracy of the sky models is much reduced relative to more typical interferometric data processing. 

\subsection{The HERA Instrument}\label{sec:HERA}

Located in the Karoo desert in South Africa at a latitude of $-30\fdg72$, HERA will consist of 350 close-packed 14~m dishes with a shortest antenna spacing of 14.6~m. Its triple split-core hexagonal layout is optimized for both redundant calibration and the delay spectrum technique for detecting the cosmic EoR \HI\ signal \cite{dil16}. 

The data used in this paper were obtained with 61 HERA dishes in total with dipole feeds spanning 100--200~MHz, of which only 50 unflagged antennas were used here. The models, however, use all 61 antennas. Although the number of antennas used in the modeling is a few higher than in the data, it will be justified later that it makes no significant difference to the comparisons or the conclusions drawn in this paper. The actual measured positions of antennas in this layout at the HERA site, including deviations from redundancy, were used in the modeling. Therefore, this accounts for a portion of the non-redundancy that may be present in the measurements. 

We used the antenna power pattern models produced by \cite{fag19} and interpolated them to the frequencies spanning the observing band, which thus far have provided a reasonably accurate characterization of the HERA antenna directivity pattern available for dish-dipole feed combination based on measurements discussed in \cite{nun20}. The power patterns of all antennas are assumed to be identical. Further, we have assumed that all antennas have an effective area of $A_\textrm{e}\simeq 100$~m$^2$ in the spectral window analyzed in this paper.   

\subsection{Foregrounds}\label{sec:FG-model}

For both fields, we have used the GaLactic and Extragalactic All-sky MWA Survey catalog \cite[GLEAM;][]{hur17} of point sources that fall within a 15\arcdeg\ radius of the telescope pointing as the sky drifts during the course of the observations. Since the GLEAM catalog does not include strong point sources, for the Fornax field, we used a separate model for the Fornax~A radio source created using clean components derived from the deconvolution techniques described in \cite{sul12} (personal communication: Carroll, Byrne 2018). Here, we have assumed that all the components of Fornax~A have a spectral index $\approx -0.78$ \cite{mck15}.

To include diffuse emission, the Galactic emission model \cite{gsm2008} was investigated which qualitatively reproduced broad spectral features ($\sim 10$~MHz scales) observed in the data \cite{car20}. However, we do not include this in the foreground model in this paper because these features are so broad that they contribute power only on the few lowest-valued $k_\parallel$-modes which are dominated by foregrounds and are not considered viable for EoR detection in this \textit{foreground avoidance} approach, and neither the Galactic emission model nor the antenna power pattern model is sufficiently precise to explain these broad oscillating spectral features in the data with high numerical accuracy \cite{car20}. 

\subsection{EoR \HI\ model}\label{sec:EoR-model}

We used a realization of a {\sc faint galaxies} EoR model \cite{mes16,gre17b} publicly available\footnote{\url{http://homepage.sns.it/mesinger/EOS.html}} from 21cmFAST simulations \cite{mes11} as our fiducial EoR \HI\ model. This model entails reionization by star forming galaxies down to low mass, normalized to match current constraints on the reionization history, and the cosmic star formation history. The lightcone cube produced by 21cmFAST spanned 1.6~Gpc ($\simeq 10$\arcdeg) in the transverse plane of the sky and redshifts extended over the 100--200~MHz passband of the instrument. This cube was smoothed and down-sampled in the transverse direction to an effective angular resolution of $\simeq 14$\arcmin~ and tiled three times on each side to produce a total angular extent of $\simeq 30$\arcdeg~ along each side in the transverse plane. The visibilities and images produced from this fiducial EoR model are presented below as well as in \cite{car20} in more detail. 

\subsection{Noise}\label{sec:Noise-model}

We have used the findings in \cite{CarilliMemo60,NimaMemo59} to guide our noise model, of the form:
\begin{equation}\label{eqn:Tsys-model}
    T_\textrm{sys}(f) = T_\textrm{rx} + T_\textrm{ant}(f_0)\left(\frac{f}{f_0}\right)^\alpha,
\end{equation}
where, $T_\textrm{sys}$ is the system temperature, $T_\textrm{rx}$ is the receiver temperature (assumed to be independent of frequency), $f_0$ is a reference frequency, and $T_\textrm{ant}$ is the antenna temperature with a spectral index $\alpha$. We have used $T_\textrm{rx}=162$~K, $f_0=150$~MHz, $T_\textrm{ant}(f_0)=200$~K, and $\alpha=-2.55$. The noise (in units of flux density) is drawn from a Gaussian distribution with standard deviation, $\sigma_T = 2 k_\textrm{B} T_\textrm{sys} / (A_\textrm{e}\sqrt{\Delta f \Delta t})$, where, $\Delta f=97656.25$~Hz and $\Delta t=10.7$~s, and is added to the simulated visibilities. By comparing to data, we will show later that this model reproduces the noise level observed in the measurements used in this work reasonably well. 

We note the noise characteristics and the beam modeling (shape and efficiency), as a function of frequency, are currently under investigation for HERA, with current estimates, for example, of $T_\textrm{sys}$ spectrum differing by up to 50\%, using different techniques \cite[see e.g.][]{CarilliMemo60, parsonsmemo34}. Moreover, the HERA feeds and electronics systems are currently being changed. Hence, modeling of the system response presented here is representative, but not final. 

\subsection{Summary of Modeling}\label{sec:modeling-summary}

\begin{figure}
\includegraphics[width=\linewidth]{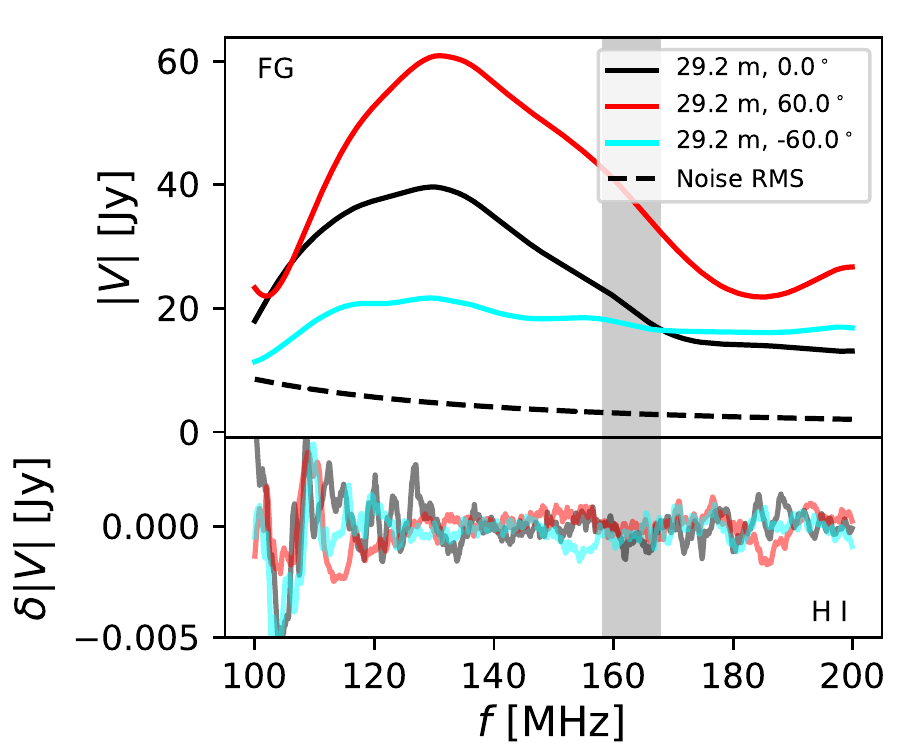}
\caption{The model of foreground visibility (top) corresponding to the J0136-30 field and the changes to the net amplitude (bottom) caused by the EoR \HI\ fluctuations (using a fiducial 21cmFAST EoR model) on a 29.2~m antenna spacing in three different orientations: 0\arcdeg (black), 60\arcdeg (red), and $-60$\arcdeg (cyan). The fluctuations in the visibility amplitude caused by the EoR \HI\ signal are noted to be a factor $\lesssim 10^{-4}$ relative to the foregrounds. The dashed line shows the spectrum of thermal noise \textit{rms} corresponding to $T_\textrm{sys}$ in units of flux density with temporal and spectral averaging over 10.7~s and 97.65625~kHz respectively. The gray shaded region shows the 163~MHz ($z\simeq 7.7$) sub-band of effective bandwidth $\Delta B=10$~MHz considered in this work. A color version of this figure is available in the online journal. 
\label{fig:vis-29m-RA0136}}
\end{figure}

Figure~\ref{fig:vis-29m-RA0136} shows the modeled visibilities from the foreground radio sources in the J0136-30 field on the 29.2~m antenna spacing in three different orientations (0\arcdeg, 60\arcdeg, and $-60$\arcdeg), and the change in the net amplitude caused by the fiducial model of EoR \HI\ fluctuations on these antenna spacings. The amplitude fluctuations are noted to be a factor $\lesssim 10^{-4}$ relative to the net visibility for the J0136-30 field. Also, shown is the spectrum of thermal noise standard deviation corresponding to the modeled $T_\textrm{sys}(f)$ in units of flux density. We consider a spectral window, $W(f)$ that is a modified Blackman-Harris window function \cite{thy16} centered at 163~MHz ($z\simeq 7.7$). $W(f)$ is defined over the entire bandpass but is shaped to have a net effective bandwidth, $\Delta B=10$~MHz.

\begin{figure}
\includegraphics[width=\linewidth]{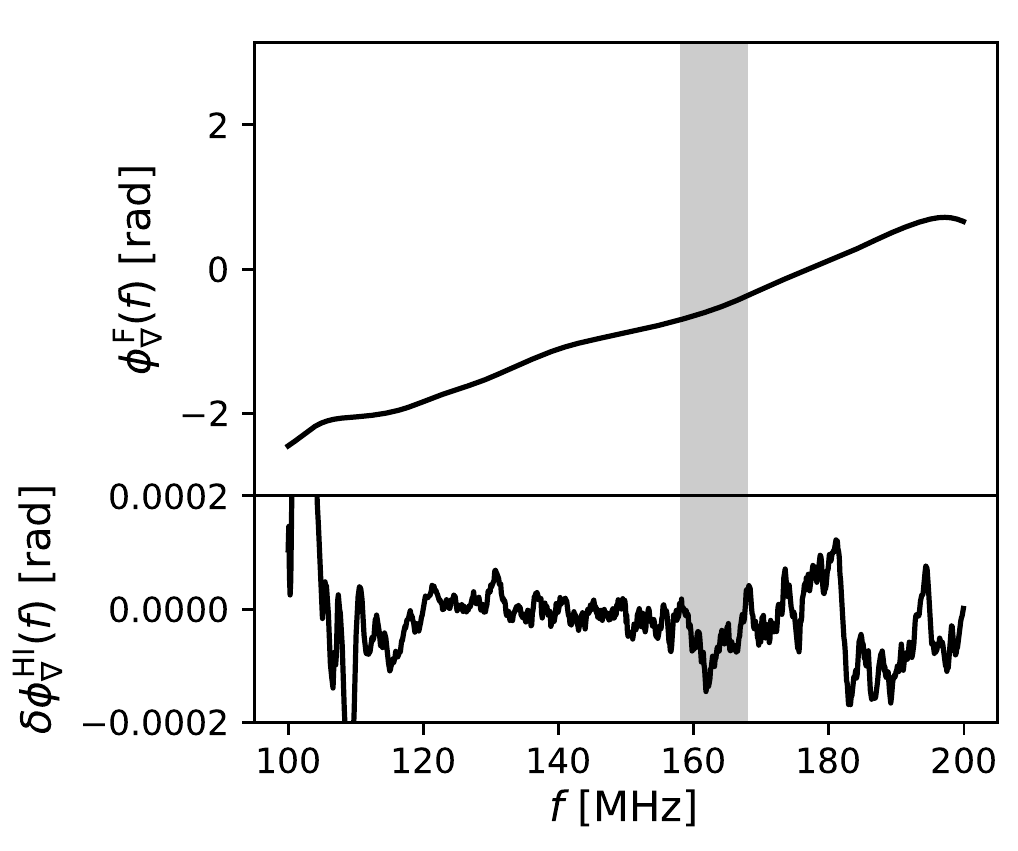}
\caption{The model of bispectrum phase angle (in radians) corresponding to the J0136-30 field on an equilateral 29.2~m antenna triad for the foregrounds (top) and the fluctuations in this phase angle caused by the EoR \HI\ fluctuations (bottom) using a fiducial 21cmFAST EoR model. The fluctuations in the bispectrum phase angle caused by the EoR \HI\ signal are noted to be $\lesssim 10^{-4}$~radians, which is comparable to the ratio of the strength of the EoR fluctuations relative to the foregrounds (see Fig.~\ref{fig:vis-29m-RA0136}). The gray shaded region shows the 163~MHz ($z\simeq 7.7$) sub-band of effective bandwidth $\Delta B=10$~MHz.
\label{fig:cPhase-EQ29m-RA0136}}
\end{figure}

Figure~\ref{fig:cPhase-EQ29m-RA0136} shows the modeled bispectrum phase angle spectrum in the J0136-30 field expected from the foregrounds and the fluctuations therein due to the EoR \HI\ fluctuations obtained using the fiducial 21cmFAST EoR model. The phase angle fluctuations are typically $\lesssim 10^{-4}$~radians, agreeing with the ratio expected between the strength of the EoR fluctuations and the foregrounds. It is these spectral fluctuations in bispectrum phase angle that we aim to detect using this new approach.  

\subsection{A Reference Delay Power Spectrum Model}\label{sec:ref-delay-power-spectrum}

As a convenient reference for comparison, using the same parameters that we adopted for the main analysis using bispectrum phase, we produce different components of a standard delay power spectrum for the simple case of \textit{foreground avoidance} on the J0136-30 field using the models described above. 

\begin{figure}
\includegraphics[width=\linewidth]{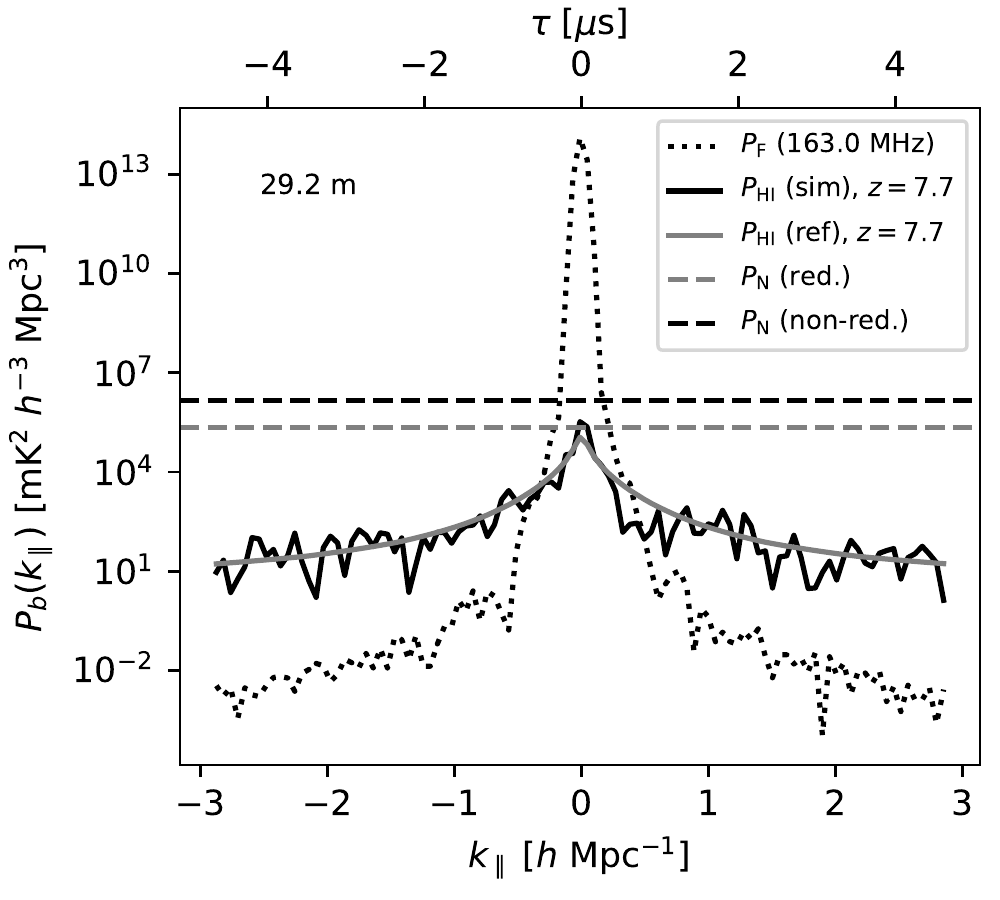}
\caption{The standard delay power spectrum of foreground and \HI\ models for the J0136-30 field in the $z=7.7$ sub-band for a 29.2~m antenna spacing. The black dotted line denotes the foreground contribution to the delay power spectrum. The black solid line denotes the EoR \HI\ power spectrum contribution from the simulation in this work including instrumental effects. The gray solid line denotes the reference EoR \HI\ power spectrum from 21cmFAST without including any instrumental effects. We note that the simulated \HI\ delay power spectra agree with the reference EoR \HI\ power spectra from 21cmFAST quite well. The unsmooth structures in the simulated delay power spectra are potentially attributable to cosmic variance since only one field was used in the simulation. The black and gray dashed horizontal lines denote the thermal noise \textit{rms} in the power spectrum, $P_\textrm{N}$, for incoherent (antenna spacings cannot be assumed to be redundant) and coherent (antenna spacings contain redundancy) averaging, respectively, of the visibilities on the different 29.2~m antenna spacings in the HERA layout considered in this work.  
\label{fig:delay-PS-29m-J0136-30-z7}}
\end{figure}

Figure~\ref{fig:delay-PS-29m-J0136-30-z7} shows the standard delay power spectrum of the foreground sky, the EoR \HI\ fluctuations, and thermal noise in the J0136-30 field on a 29.2~m antenna spacing in a sub-band corresponding to $z=7.7$. The reference input \HI\ power spectrum from 21cmFAST (solid gray) and the simulated EoR \HI\ power spectrum after including the instrumental effects through PRISim (solid black) agree well with each other. This further validates the simulation as was also verified in \cite{car20}. The minor unsmooth structures in the simulated EoR \HI\ power spectrum are potentially attributable to cosmic variance since only one realization of the \HI\ fluctuations as observed on one field was simulated \cite{lan19}. The foreground power (dotted lines) dominate over the cosmological \HI\ signal from the EoR on $|k_\parallel| \lesssim 0.4\,h$~Mpc$^{-1}$. This indicates that the EoR signal will be separable from the foregrounds using a purely \textit{foreground avoidance} technique only at $|k_\parallel| \gtrsim 0.4\,h$~Mpc$^{-1}$ for the 29.2~m antenna spacings.

The thermal noise \textit{rms} in the power spectra, $P_\textrm{N}$ (see \S\ref{sec:analytical-thermal-noise-power} for details) is shown for the two cases when redundancy of visibilities on nominally redundant antenna spacings can and cannot be assumed in dashed gray and dashed black lines respectively, corresponding to the data volume, observing parameters, and various processing steps. It must be noted that the thermal noise power \textit{rms} was derived quasi-analytically and should be considered as ideal and the best-case scenario.

\section{Analysis}\label{sec:Analysis}

While describing the analysis steps below of the bispectrum phase, we note that they are applied identically to the data and the forward-modeling.  

\subsection{Data Selection}\label{sec:Data-selection}

Throughout this analysis, we work with raw, uncalibrated data unless otherwise indicated. We restrict our data selection to equilateral triads of 29.2~m antenna spacings. The equilateral shape allows an easier interpretation of the $k_\perp$ modes corresponding to these antenna spacings. The raw data from antenna spacings that comprise these triads are combined to obtain the bispectrum phase. At this stage, the data array has dimensions $N_l^\prime \times N_r^\prime \times N_\nabla \times N_f$, where, $N_l^\prime$ denotes the number of time intervals contiguous in \textit{local sidereal time} (LST) with a temporal resolution of 10.7~s, $N_r^\prime = 18$ denotes the number of repeated nights spanning the same LST range, $N_f = 1024$ is the number of frequency channels, and $N_\nabla = 31$ is the number of 29.2~m equilateral triads in the data (from 50 unflagged antennas only) and 37 in the models (from all 61 antennas). We consider all 37 triads in the model to estimate the best-case hypothetical sensitivity expected from using all available antennas in the current layout.

For this first demonstration, we explored the data for spectral windows with minimal Radio Frequency Interference (RFI) and found one centered on 163~MHz corresponding to $z\simeq 7.7$ (shaded band in Fig.~\ref{fig:vis-29m-RA0136}). We neither flag nor correct these data subsets explicitly for RFI and therefore, they may still contain RFI at low levels. 

We denote the bispectrum phase angle as $\phi_\nabla(f)$, which is obtained from 
\begin{align}
    e^{i\phi_\nabla(f)} &= \prod_{p=1}^3\frac{ V_p(f)}{|V_p(f)|},
\end{align}
where, $V_p(f)$ denotes the visibility spectrum measured on antenna spacings, $\boldsymbol{b}_p$, indexed by $p=\{1,2,3\}$, comprising a triad denoted by $\nabla$.

\subsection{Coherent Averaging}\label{sec:Coherent-averaging}

Coherent averaging of the bispectrum phase can significantly improve sensitivity to the final power spectrum. This requires either an assumption or the knowledge that the underlying quantity of scientific interest being averaged, such as the sky-based signal, does not vary significantly along the axis/axes being averaged, otherwise it may lead to loss of sensitivity as the signal strength may not be preserved in the averaging process. 

We note that by `average', we refer to both the mean and median metrics. The latter is specially useful in rejecting outliers, such as RFI-affected subsets of data, and minimizing their effect on further processing and the final power spectrum estimates. Other than the usage of the median statistic, no special flagging or mitigation of RFI is done. 

Based on Allan-variance analysis in \cite{car18}, the bispectrum phases can be `averaged' to 1~min intervals along LST while still improving sensitivity. We have limited our LST range to $\sim 22$~min centered on the transit of the two fields, namely, Fornax~A and J0136-30. Because HERA is a redundant array, it offers another dimension along which coherent `averaging' can be performed, namely, the redundant triads that fall under the class of 29.2~m equilateral triads. The nature and quality of redundancy of the HERA measurements is currently under investigation \cite{ken19}. Therefore, in this paper we take two approaches -- a conservative approach by not making any assumption that the measured bispectrum phases on nominally redundant triads are actually redundant but statistically they share common traits, such as the underlying power spectrum; and, an approach that relies on exploiting the assumed redundancy in the nominally redundant triads.  

We have observed that the bispectrum phases repeat accurately across different nights at a given LST, and can thus be `averaged' together coherently in phase \cite{car18,car20}. Further, in order to obtain power spectrum estimates free of noise bias, we do not average all the repeated nights together at once, but instead divide them into two equal-sized bins containing 9 nights each, which then get averaged separately. These two subsets will later be used in cross-power spectrum estimation. While this process does not achieve maximum sensitivity because of incomplete coherent `averaging', it offers the benefit of eliminating noise-bias in the power spectrum since the noise is uncorrelated between the two subsets. After coherent reduction, the dimensions of the array of measurements are $N_l \times N_r \times N_\nabla \times N_f$, where, $N_l = 22$, and $N_r = 2$.    

\subsection{Spectral Windowing, Scaling to Flux Density, and Delay Spectrum}\label{sec:DSpec}

The spectral window function, $W(f)$, can be used to control the quality of the delay spectrum \cite{thy13,thy16}. We use a modified Blackman-Harris window function to obtain substantial gain in dynamic range while accepting a small compromise on the resolution in Fourier space \cite{thy16}. This ensures that the power spectra presented here are not limited by the dynamic range of the spectral window function. Although $W(f)$ is defined over the entire bandpass, to minimize cosmic signal evolution, we shape it such that it has an effective bandwidth, $\Delta B\simeq 10$~MHz. 

The next step in the analysis involves applying a scaling factor to the bispectrum phases. The motivation and its derivation are described in detail in Paper~I, with a brief summary here. The bispectrum phase angle fluctuations due to \HI\ from the EoR, when small relative to the foreground strength, have been shown in Paper~I to be:
\begin{align} \label{eqn:bispectrum-line-perturbations}
  \delta\phi_\nabla^\textrm{H{\sc i}}(f) &\approx -i\,e^{i\delta\phi_\nabla^\textrm{H{\sc i}}(f)} \approx \sum_{p=1}^3 \, \Im\left\{\frac{V_p^\textrm{H{\sc i}}(f)}{V_p^\textrm{F}(f)}\right\},
\end{align}
where, $V_p^\textrm{H{\sc i}}(f)$ and $V_p^\textrm{F}(f)$ denote the visibilities due to the EoR \HI\ and the foregrounds, respectively, and $\Im\{\cdot\}$ denotes the imaginary part. If a model for $|V_p^\textrm{F}(f)|$ is available, it can be used to extract partial information about $V_p^\textrm{H{\sc i}}(f)$. Let $V_\textrm{eff}^\textrm{F}$ denote such an empirical model of the true sky-based foreground visibility averaged over the sub-band derived from estimates of $V_p^\textrm{F}(f)$.

We apply the same spectral window function on the visibility models, either from data or from simulations, that have been already calibrated and averaged over repeated nights and redundant antenna spacings. $V_\textrm{eff}^\textrm{F}$ is then obtained by estimating the mean of the visibilities weighted in inverse-quadrature within the sub-band as a scalar as a function of antenna spacings, and LST. $V_\textrm{eff}^\textrm{F}$ does not have any spectral dependence and serves as a single scaling factor, appropriate to the sub-band, primarily to remove the modulating effect of the strength of the foregrounds in Eq.~(\ref{eqn:bispectrum-line-perturbations}), besides converting the complex exponentials of the binned and `averaged' bispectrum phases to have units of flux density.

For any given triad, there are three such visibilities. However, only one scaling factor is applied. Hence, there is a choice to be made about the combination of the three visibilities that yields this scaling factor. The noise in the bispectrum phase angle is the sum of the three visibility phase angle noises and is thus predominantly determined by the visibility which contains the least foreground amplitude assuming equal noise \textit{rms} on all three visibilities. In this work, we obtain the flux density scaling factor by summing the absolute values of the three visibilities in inverse quadrature. The binned and `averaged' bipectrum phases are converted to flux density units by multiplying with this absolute visibility amplitude. It must be noted that calibrated visibilities are only used toward the purpose of obtaining a single scaling factor over the entire sub-band and thus do not introduce spectral errors due to the lack of accurate calibration. Thus, an estimate of the measured (calibrated) visibility can be reconstructed as:
\begin{align}
    V_\nabla(f) &= V_\textrm{eff}^\textrm{F} \, e^{i\phi_\nabla(f)}.
\end{align}

We obtain the delay spectrum \cite{par12b} of the binned, `averaged', and scaled bispectrum phase using:
\begin{equation}\label{eqn:cpdspec}
  \widetilde{\Psi}_\nabla(\tau) = V_\textrm{eff}^\textrm{F}\,\int e^{i\phi_\nabla(f)}\,W(f)\,e^{i2\pi f\tau}\,\mathrm{d}f.
\end{equation}
$\widetilde{\Psi}_\nabla(\tau)$ is downsampled along the $\tau$ axis to contain only independent samples with a resolution $\delta\tau=1/\Delta B$. At this stage, the delay spectrum products have dimensions $N_z \times N_l \times N_r \times N_\nabla \times N_\tau$, where, $N_\tau = 103$ (after downsampling), and $N_z$ denotes the number of spectral sub-bands being processed ($N_z=1$ corresponding to the $z=7.7$ sub-band in this paper). 

\subsection{Cross-Power Spectra}\label{sec:PSpec}

We define the delay cross-power spectrum from two independent realizations of delay spectra, $\widetilde{\Psi}_\nabla(\tau)$ and $\widetilde{\Psi}_\nabla^\prime(\tau)$, assuming the EoR \HI\ signal is coherent between the two, as:
\begin{align}
P_\nabla(\kappa_\parallel) &\equiv \Re\bigl\{\widetilde{\Psi}_\nabla(\tau)\,\conj{\widetilde{\Psi}^\prime_\nabla}(\tau)\bigr\} \nonumber\\ 
&\qquad \times\left(\frac{A_\textrm{e}}{\lambda^2\Delta B}\right) \left(\frac{D^2\Delta D}{\Delta B}\right)\left(\frac{\lambda^2}{2k_\textrm{B}}\right)^2, \label{eqn:cross-pspec}
\end{align}
with
\begin{align}
  \kappa_\parallel &\equiv \frac{2\pi\tau\,f_\textrm{r}H_0\,E(z)}{c(1+z)^2}, \label{eqn:tau-kprll}
\end{align}
where, $\Re\{\cdot\}$ denotes the real part, $\conj{Z}$ denotes the complex conjugate of $Z$, $\lambda$ is the wavelength of the band center, $D\equiv D(z)$ is the comoving distance to redshift $z$, $\Delta D$ is the comoving depth along the line of sight corresponding to $\Delta B$, $k_\textrm{B}$ is the Boltzmann constant, $c$ is the speed of light in vacuum, $f_\textrm{r}$ is the rest-frame frequency of the electronic spin-flip transition of \HI\, and $h$, $H_0$ and $E(z)\equiv [\Omega_\textrm{M}(1+z)^3+\Omega_\textrm{k}(1+z)^2+\Omega_\Lambda]^{1/2}$ are standard terms in cosmology.

$\kappa_\parallel$ and $P_\nabla(\kappa_\parallel)$ defined here have units of $h\,$Mpc$^{-1}$ and $\textrm{mK}^2\,(\textrm{Mpc}/h)^3$ respectively. However, because the origin of the EoR \HI\ fluctuations in bispectrum phase are not the same as those in the usual scenario with visibilities (Paper I), we caution that the $\kappa_\parallel$ and $P_\nabla(\kappa_\parallel)$ used here in the context of the bispectrum phase are not to be interpreted physically, but only as mathematical analogs to the standard terms used in cosmology. Hereafter, the units of $\kappa_\parallel$ and $P_\nabla(\kappa_\parallel)$ will be designated by ``pseudo $h\,$Mpc$^{-1}$'' and ``pseudo $\textrm{mK}^2\,(\textrm{Mpc}/h)^3$'' respectively.

The full cross-product array, $P_\nabla(\kappa_\parallel)$, will have dimensions $N_z \times N_l \times N_l \times N_r \times N_r \times N_\nabla \times N_\nabla \times N_\tau$. In order to minimize memory requirements while computing, we do not form every pair of cross-power product possible. Instead, we only form the averages across the main and off-diagonals of these square matrices of cross-products, which are incoherent in nature because they were obtained by averaging cross-power spectra. The diagonally averaged cross-power spectra have dimensions $N_z \times N_{\delta l} \times N_{\delta r} \times N_{\delta\nabla} \times N_\tau$, where, the difference in indices denotes the offset from the main diagonal. Therefore, $\delta l\in [-(N_l-1), (N_l-1)]$, $\delta r \in [-(N_r-1), (N_r-1)]$, and $\delta\nabla \in [-(N_\nabla-1), (N_\nabla-1)]$. It may be noted that the lower triangular portions of the full cross-power data will be complex conjugates of the upper triangular positions and therefore do not contain independent information. Hence, only the non-negatively offset diagonals, including the main diagonal, are retained for further analysis.

\subsection{Incoherent Averaging of Cross-Power Spectra}\label{sec:Incoherent-averaging}

In principle, we could assume that the data are redundant between bins across repeated nights or across redundant triads or both.  In this paper, we consider either of these cases when such an assumption may or may not be valid. However, in either case, an assumption of redundancy and zero signal de-correlation will only be valid for $\delta l = 0$. Therefore, in subsequent analysis, we choose cross-power spectra with $\delta l=0$, $\delta r = 1$, and $\delta\nabla \ge 0$, where $\delta r = 1$ or $\delta\nabla > 0$ are expected to provide estimates free of noise-bias. 

\subsection{Estimating Uncertainties in Power Spectra}\label{sec:Uncertainties}

We divide the raw bispectrum phases into roughly four equal bins across the repeated nights of observing and average inside the respective bins. This allows us to choose a pair of bins and calculate the difference. With four bins, three unique non-repeated pairs of bins are available for differencing. It must be noted that since the sky contribution to bispectrum phases are found to repeat across nights at a given LST, these differences will have eliminated any sky contribution and they will be comprised purely of systematics and noise realizations that are not common across these bins. The cross-power between these three difference-pairs will produce three estimates of the uncertainty on the power spectrum from which the average uncertainty can be estimated. 

This empirical estimate of the uncertainty has the advantage of including both the thermal noise and the systematic components such as from RFI, non-redundancy, LST mis-alignment, etc. An estimate of the noise-only component can be obtained from using the model and comparing it with the data.

This method has some potential disadvantages. Firstly, a power spectrum in reality will have uncertainties that comprise of cross-terms between the sky signal and the thermal noise contributions, in addition to the noise-only cross-power. The uncertainty due to the sky-and-noise cross-power will be foreground-dominated in the lowest-valued spectral modes. This method will not contain this contribution since any sky contribution was eliminated in the initial differencing process and thus may underestimate the actual uncertainty in the lowest-valued spectral modes. The higher-valued spectral modes, however, will remain unaffected as they are expected to be predominantly dominated by thermal noise alone and hence, will be not be underestimated. Secondly, since the average level of uncertainty is determined from only three pairs of subsets, the average uncertainty so estimated will itself be subject to a significant level of uncertainty.

Although the information about the power spectrum of the sky component including the foregrounds and the EoR signal is nominally contained in the real part (see Eq.~(\ref{eqn:cross-pspec})), the presence of noise will introduce an imaginary component in the cross-power spectrum. Similarly, any systematics in the instrument or the analysis, or non-stationarity in the sky signal will also contribute to the imaginary component. The excess over the thermal noise component in the imaginary part is a measure of the systematics present in the data and the uncertainties therein. In order to provide the overall uncertainty including all these effects and not underestimate our net uncertainty, we conservatively include the imaginary component in addition to the real component to derive our standard deviation that will be shown in the results that follow. It must be noted that these errors derived as described above are not expected to follow a Gaussian distribution even in the ideal case where systematic effects are absent. Therefore, we do not make any assumption about the probability density function of these errors and use the standard $D_\textrm{KS}$ and $P_\textrm{KS}$ metrics of the non-parametric two-sample Kolmogorov-Smirnov test \cite[K-S test;][]{hod58} to evaluate the errors in the data and compare them statistically with the models using a significance threshold equivalent to $5\sigma$ rejection in a Gaussian distribution.

\section{Results}\label{sec:Results}

In this section, we show the results of our analysis on the Fornax and J0136-30 fields. We consider two different cases free of noise-bias -- the conservative case is the average of the `intra-triad' cross-power obtained when $\delta\nabla=0$ (main diagonal) where redundancy has not been assumed even on nominally redundant triads, and the other case is the average `inter-triad' cross-power obtained when $\delta\nabla > 0$ (upper triangle) while assuming and then exploiting the  redundancy between nominally redundant triads. 

In the discussions hereafter, we highlight the following distinction between the usage of the terms ``noise-limited'' and ``data-limited''. The former signifies that the uncertainties are at a level expected from thermal noise contributions alone and are ``noise-like'', whereas the latter signifies that the uncertainties may be larger than the contributions from thermal noise alone due to contributions from systematics and other effects but are still randomly behaved and are capable of being mitigated further by using more data.

\subsection{Intra-triad Cross-Power Spectrum}\label{sec:Intra-triad-PSpec}

\begin{figure*}
\centering
\subfloat[][Data\label{fig:EQ28XX_data_Fornax_z_7.7_triadcomb_0}]{\includegraphics[width=0.47\textwidth]{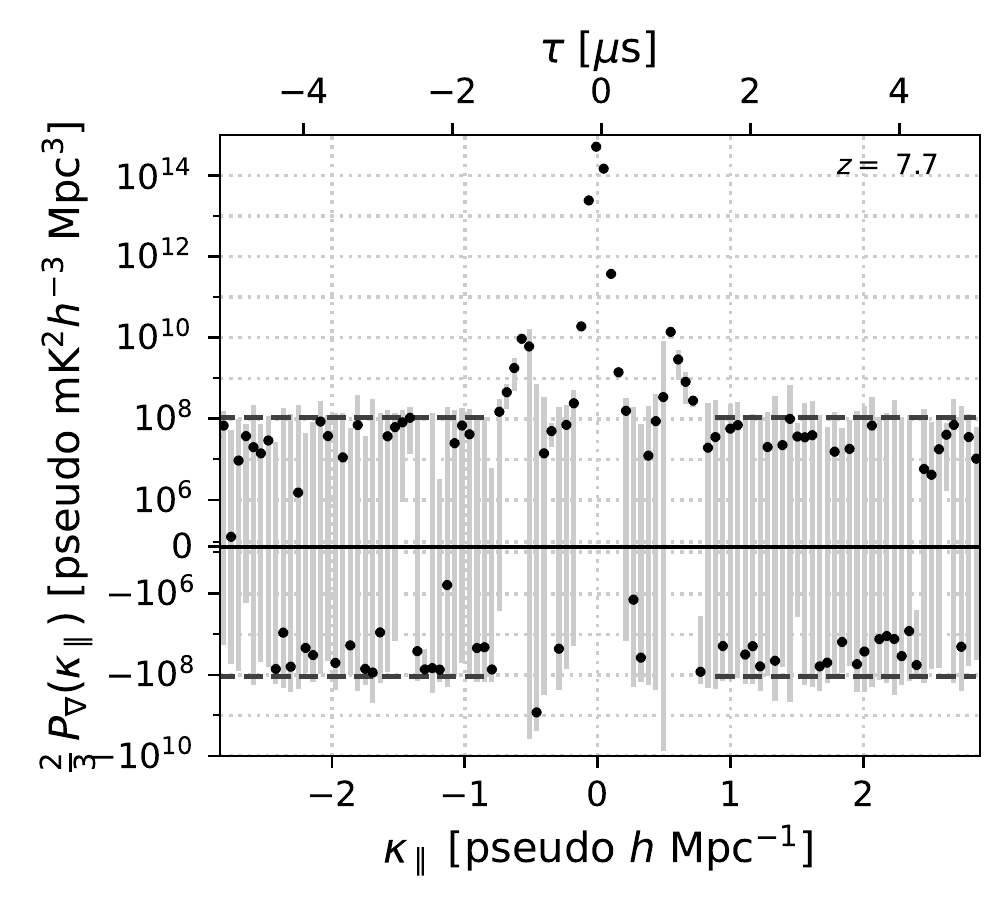}}
% \subfloat[][Data\label{fig:EQ28XX_data_Fornax_z_7.7_triadcomb_0}]{\includegraphics[width=0.47\textwidth]{figures/EQ28XX_data_Fornax_transit_cpdps_z_7_median_resampled_dlst_60s_triadcomb_0.pdf}}
\subfloat[][Model\label{fig:EQ28XX_model_Fornax_z_7.7_triadcomb_0}]{\includegraphics[width=0.47\textwidth]{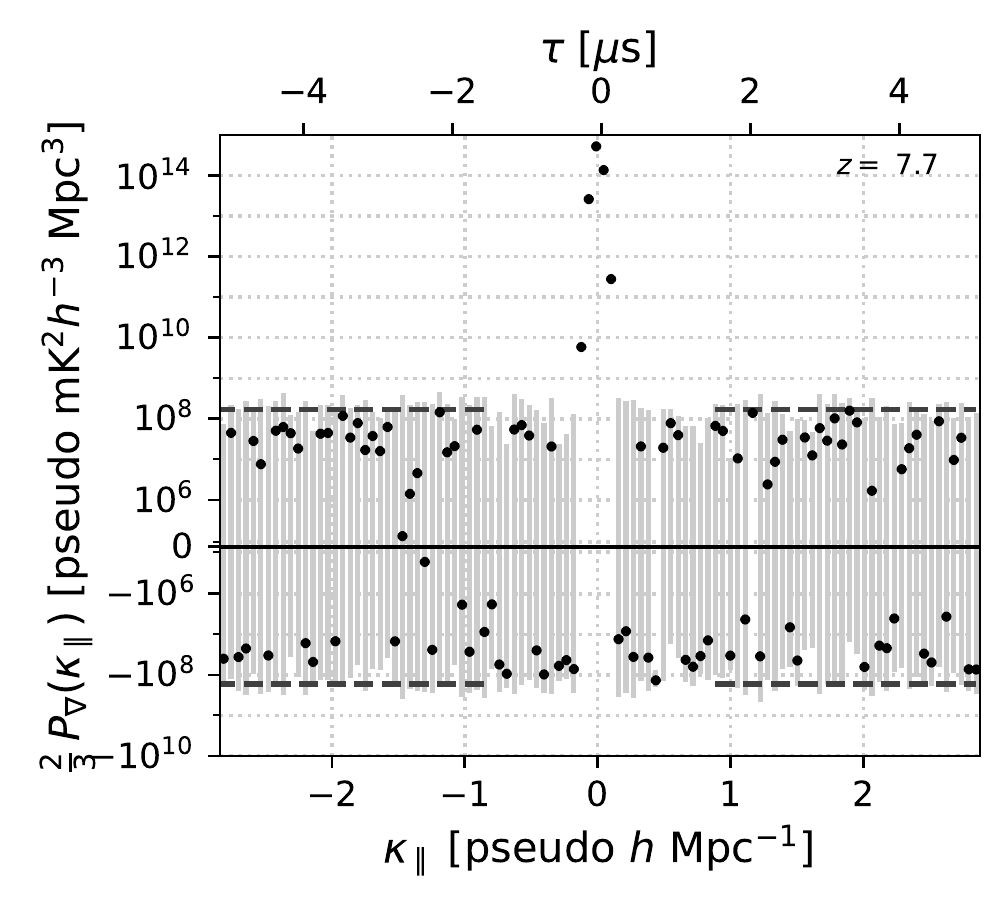}}
% \subfloat[][Model\label{fig:EQ28XX_model_Fornax_z_7.7_triadcomb_0}]{\includegraphics[width=0.47\textwidth]{figures/EQ28XX_model_Fornax_transit_cpdps_z_7_median_resampled_dlst_60s_triadcomb_0.pdf}}
\caption{Two-thirds of the delay power spectrum of the bispectrum phase in Eq.~(\ref{eqn:cross-pspec}) obtained by averaging the power spectra from individual triads (intra-triad, $\delta\nabla=0$) for the Fornax field from the HERA data (left) and corresponding modeling (right) at a frequency sub-band centered at $z=7.7$ weighted by a modified Blackman-Harris spectral window function of effective bandwidth 10~MHz. The $x$-axis on the top corresponds to $\tau$ linearly related to $\kappa_\parallel$ by Eq.~(\ref{eqn:tau-kprll}). The black symbols denote the data (left) and the matching model comprising of foregrounds and noise (right). The $y$-axis is shown on a symmetric logarithmic scale, where the solid horizontal line denotes the zero-point below which the values of the power spectrum are negative. The gray vertical lines denote uncertainties (one standard deviation or ``$1\sigma$'', but not necessarily associable with a Gaussian distribution) estimated by splitting the data into four roughly equal parts spanning the 18 nights of repeated observations but have independent noise and estimating the power from the differences between these subsets. The horizontal dashed lines denote the standard deviation of the bin-to-bin variation in $|\kappa_\parallel| \gtrsim 0.85\,\textrm{``pseudo''}\,h$~Mpc$^{-1}$, which for the data and the model are $\simeq 1.1\times 10^8$~``pseudo''~mK$^2(\textrm{Mpc}/h)^3$ and $\simeq 1.7\times 10^8$~``pseudo''~mK$^2(\textrm{Mpc}/h)^3$ respectively. In a majority of $\kappa_\parallel$-bins ($|\kappa_\parallel| \gtrsim 0.85\,\textrm{``pseudo''}\,h$~Mpc$^{-1}$), these uncertainties include the zero-point. A two-sided K-S test does not reject that the null hypothesis that errors in the data and the model could have been drawn from the same distribution at a high significance. This supports that the power spectrum estimates in these $\kappa_\parallel$-bins are predominantly noise-limited. Both the data and the model have a central peak at $\kappa_\parallel=0\,\textrm{``pseudo''}\,h$~Mpc$^{-1}$ and a noise floor at $|\kappa_\parallel| \gtrsim 0.85\,\textrm{``pseudo''}\,h$~Mpc$^{-1}$ consistent with the model of foregrounds including noise (right). The data exhibit secondary peaks at $\kappa_\parallel \simeq 0.5\,\textrm{``pseudo''}\,h$~Mpc$^{-1}$, which have not been modeled in the right panel and are believed to be caused by baseline-dependent errors, such as cross-talk between pairs of antennas \cite{ker20a}. The values around these secondary peaks appear to be systematic-limited contrasting the rest of the bins where they are predominantly noise-limited. 
\label{fig:EQ28XX-baseline-cpdps-Fornax-z-7-median}}
\end{figure*}

Figure~\ref{fig:EQ28XX-baseline-cpdps-Fornax-z-7-median} shows two-thirds of the intra-triad cross-power spectrum of the bispectrum phase obtained on the Fornax field in the $z=7.7$ sub-band. The factor $2/3$ is used to normalize the power spectrum while accounting for three antenna spacings contributing to the power and only one half of the power being retained by considering only the phase of the bispectrum. The black symbols denote the power spectrum obtained with data (left) and the modeling that includes noise (right). The scaling of $y$-axis to the physical units specified is described in \S\ref{sec:DSpec}, which builds on the methodology in Paper I. The vertical gray lines denote the uncertainty ranges in the data and the noisy model obtained using the method described in Section~\ref{sec:Uncertainties}, and are not expected to follow a Gaussian distribution.

The following features are prominently noted:
\begin{enumerate}
    \item The peaks of $\simeq 10^{15}$~``pseudo''~mK$^2(\textrm{Mpc}/h)^3$ in both the data and the models at $\kappa_\parallel=0\,\textrm{``pseudo''}\,h$~Mpc$^{-1}$ are set by the foreground contribution to the bispectrum phase, which in this case is predominantly due to the strong emission from Fornax~A present near the edge of the primary beam of the antenna power pattern. Fornax~A is essentially a point source at the angular  resolution of the current HERA layout, and hence the integrated flux density is coherent on all antenna spacings. 
    \item Secondary peaks of $\simeq 10^{10}$~``pseudo''~mK$^2(\textrm{Mpc}/h)^3$ are seen at $\kappa_\parallel \simeq  \pm 0.5\,\textrm{``pseudo''}h$~Mpc$^{-1}$ in the data, corresponding to a delay, $\tau\simeq 1\,\mu$s (see Eq.~\ref{eqn:tau-kprll}). These secondary peaks are not seen in the models. The secondary peaks do not always show uncertainty ranges consistent with noise, and are therefore probably limited by systematics. The secondary peaks in the power spectrum arise due to a $\sim 1$~MHz spectral ripple that can be seen in the visibility amplitude, phase, and closure phase spectra at a level of $\sim 0.5$\%, in the worst cases \cite{car20}. References \cite{ker19,ker20a} have also observed and modeled a very similar systematic effect in their data and they conclude that it is not an antenna-based but a baseline-dependent systematic such as cross-talk between antennas. The fact that bispectrum phases are immune to direction-independent and multiplicative antenna-based systematics therefore confirms that these secondary peaks probably arise from baseline-dependent effects which could be some combination of inter-feed or inter-dish cross-coupling along over-the-air or electrical pathways. 
    \item At $|\kappa_\parallel|\gtrsim 0.85\,\textrm{``pseudo''}\,h$~Mpc$^{-1}$, both data and model have power spectra (solid circles) at the level of $\simeq 10^7$--$10^8$~``pseudo''~mK$^2(\textrm{Mpc}/h)^3$. 
    \item Avoiding the lowest-valued spectral modes where the errors may be underestimated, the \textit{rms} of the bin-to-bin variation in $|\kappa_\parallel| \ge 0.85\,\textrm{``pseudo''}\,h$~Mpc$^{-1}$ (horizontal dashed lines) obtained using a standard deviation robust to outliers is $\simeq 1.1\times 10^8$~``pseudo''~mK$^2(\textrm{Mpc}/h)^3$ and $\simeq 1.7\times 10^8$~``pseudo''~mK$^2(\textrm{Mpc}/h)^3$ in the data and the model respectively. In these spectral modes, a two-sample K-S test of the errors (gray vertical lines) yields $D_\textrm{KS}=0.075, P_\textrm{KS}=0.16$ suggesting that the null hypothesis that the samples in the model and data were drawn from the same distribution cannot be rejected with high significance.
    \item The data at $|\kappa_\parallel|\gtrsim 0.85\,\textrm{``pseudo''}\,h$~Mpc$^{-1}$ oscillate between negative and positive values roughly randomly and with equal frequency indicating a noise-limited behavior, which in turn indicates that this sub-band is relatively unaffected by systematics such as RFI.
\end{enumerate}

The above findings indicate that both the sky and noise models used in the modeling section are reasonably accurate and match this subset of the data. The dynamic range between the central peak and the noise floor is $\simeq 10^7$--$10^8$, both in the data and model. Due to strong emission from Fornax~A, the Fornax field exhibits such a high dynamic range between the central peak and the noise floor, and it could be more susceptible to limitations due to systematics as the noise floor drops with increase in sensitivity to thermal noise.

\begin{figure*}
\centering
\subfloat[][Data\label{fig:EQ28XX_data_RA1.6hr_z_7.7_triadcomb_0}]{\includegraphics[width=0.47\textwidth]{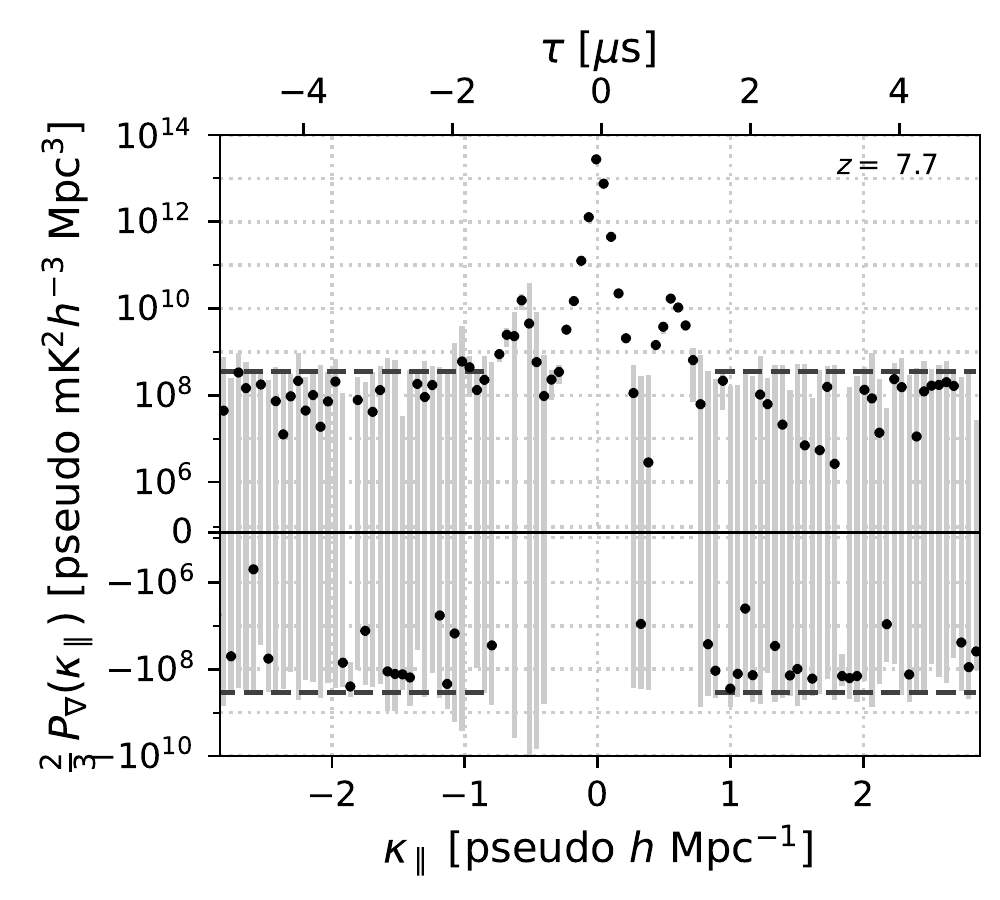}}
% \subfloat[][Data\label{fig:EQ28XX_data_RA1.6hr_z_7.7_triadcomb_0}]{\includegraphics[width=0.47\textwidth]{figures/EQ28XX_data_RA_0136_hr_transit_cpdps_z_7_median_resampled_dlst_60s_triadcomb_0.pdf}}
\subfloat[][Model\label{fig:EQ28XX_model_RA1.6hr_z_7.7_triadcomb_0}]{\includegraphics[width=0.47\textwidth]{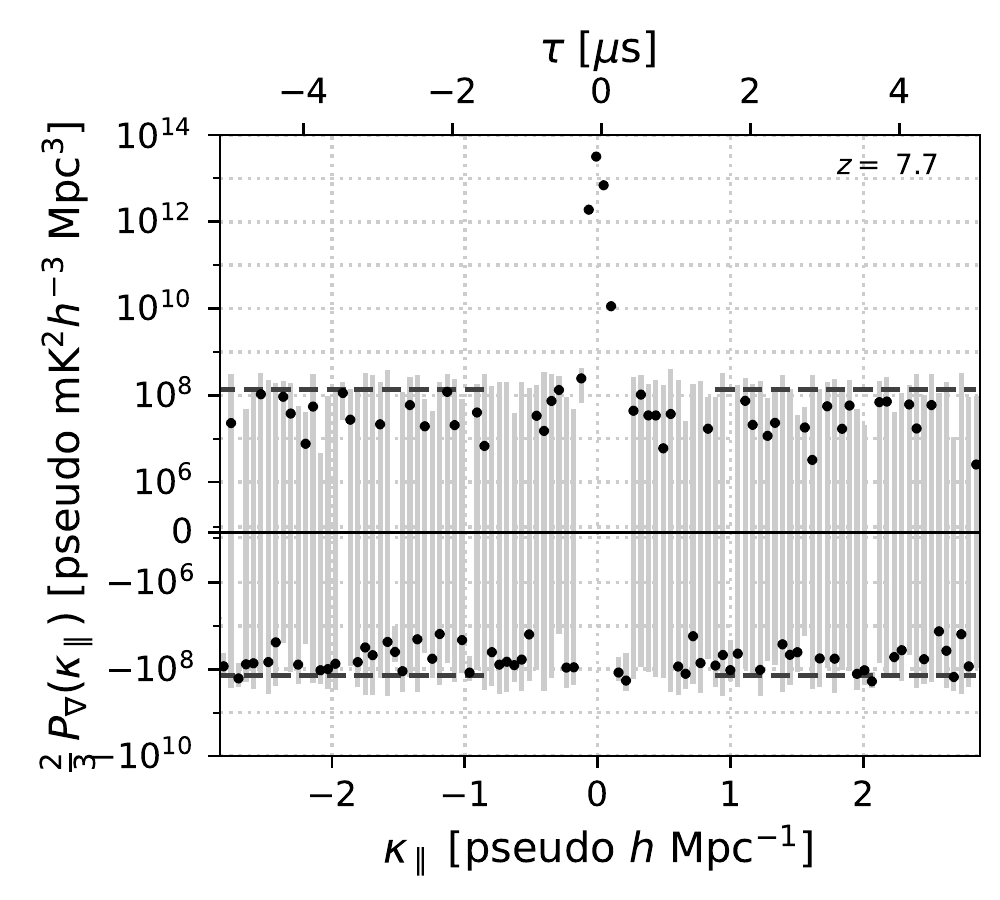}}
% \subfloat[][Model\label{fig:EQ28XX_model_RA1.6hr_z_7.7_triadcomb_0}]{\includegraphics[width=0.47\textwidth]{figures/EQ28XX_model_RA_0136_hr_transit_cpdps_z_7_median_resampled_dlst_60s_triadcomb_0.pdf}}
\caption{Same as Fig.~\ref{fig:EQ28XX-baseline-cpdps-Fornax-z-7-median} but for the J0136-30 field. The \textit{rms} of bin-to-bin variation at $|\kappa_\parallel|\gtrsim 0.85\,\textrm{``pseudo''}\,h$~Mpc$^{-1}$ is estimated to be $\simeq 3.5\times 10^8$~``pseudo''~mK$^2(\textrm{Mpc}/h)^3$ and $\simeq 1.4\times 10^8$~``pseudo''~mK$^2(\textrm{Mpc}/h)^3$ in the data and the model respectively. A hint of a bias towards positive values in the data (solid circles) is noted. Supporting this discrepancy, a two-sided K-S test shows that the errors in the data and the model are extremely unlikely to be drawn from the same distribution indicating that some systematics such as LST misalignments, RFI, or time-varying effects may be present in the data. \label{fig:EQ28XX-baseline-cpdps-RA1.6hr-z-7-median}}
\end{figure*}

Therefore, we also consider another relatively weaker field. Figure~\ref{fig:EQ28XX-baseline-cpdps-RA1.6hr-z-7-median} shows the intra-triad cross-power spectrum for the J0136-30 field.

In comparison to the Fornax field (Fig.~\ref{fig:EQ28XX-baseline-cpdps-Fornax-z-7-median}), the J0136-30 field exhibits the following characteristics:
\begin{enumerate}
    \item The peak is $\simeq 3\times 10^{13}$~``pseudo''~mK$^2(\textrm{Mpc}/h)^3$, significantly lower relative to the Fornax field because the J0136-30 field is weaker in radio emission.
    \item The secondary peaks caused by baseline-dependent systematics are seen at a level similar to the Fornax field, $\simeq 10^{10}$~``pseudo''~mK$^2(\textrm{Mpc}/h)^3$ at $|\kappa_\parallel|\simeq 0.5\,\textrm{``pseudo''}\,h$~Mpc$^{-1}$.
    \item At $|\kappa_\parallel|\gtrsim 0.85\,\textrm{``pseudo''}\,h$~Mpc$^{-1}$, the data points (solid circles) are on average slightly higher than the Fornax field whereas the model (right panel) is comparable to the Fornax field.
    \item The \textit{rms} of the bin-to-bin variation (horizontal dashed lines) at $|\kappa_\parallel|\gtrsim 0.85\,\textrm{``pseudo''}\,h$~Mpc$^{-1}$ is estmated to be $\simeq 3.5\times 10^8$~``pseudo''~mK$^2(\textrm{Mpc}/h)^3$ and $\simeq 1.4\times 10^8$~``pseudo''~mK$^2(\textrm{Mpc}/h)^3$ in the data and the model respectively. Besides this discrepancy, the presence of a visual statistical bias towards positive values and larger uncertainties in the data is noted.
    \item Consistent with the above findings, the non-parametric two-sample K-S test between the errors in the data and model yielded $D_\textrm{KS}\approx 0.23, P_\textrm{KS}\simeq 10^{-10}$ which indicates that the null hypothesis that they are drawn from the same distribution could be rejected with high significance. In other words, they are very different statistically. All these indicate that some systematic effects are significantly affecting the power spectrum of the data. The higher uncertainties in the data imply that the differenced bispectrum phases and power spectra are not noise-limited. This could be due to LST-misalignment, RFI, or other time-varying systematics that affect the differenced bins unequally. A more thorough diagnosis to understand this behavior will be undertaken but is beyond the scope of this work.  
\end{enumerate}

Since the model has more triads (37) than in the data (31), the uncertainty estimates from the models are likely to be $\approx 91.5$\% ($\approx\sqrt{31/37}$) of the uncertainty when the same number of triads in the data are used. We note that this change denotes an insignificant correction to the estimated uncertainties.

\subsection{Inter-triad Cross-Power Spectrum}\label{sec:Inter-triad-PSpec}

In order to reduce the secondary peaks at $\kappa_\parallel \simeq \pm 0.5\,\textrm{``pseudo''}\,h$~Mpc$^{-1}$ and the significant predominance of systematics in the uncertainties, we compute the cross-power spectrum between triads falling under the same class, referred to in this paper as the inter-triad cross-power spectrum. The inherent assumption here is that the triads behave redundantly towards the sky-based signal but the phase offsets of the $\sim 1$~MHz ripple and other systematics are randomly behaved across the triads. If the assumption about redundancy of nominally redundant triads to the sky signal is not valid, then there will be loss of power measured from the \HI\ signal, which eventually may or may not be scale-dependent. And if the assumption about randomness of the systematic effects across the different triads is invalid, then the power spectrum will not show any significant mitigation of the systematics observed previously in the intra-triad averaging (see Fig.~\ref{fig:EQ28XX-baseline-cpdps-RA1.6hr-z-7-median}).

\begin{figure*}
\centering
\subfloat[][Data\label{fig:EQ28XX_data_RA1.6hr_z_7.7_triadcomb_2}]{\includegraphics[width=0.47\textwidth]{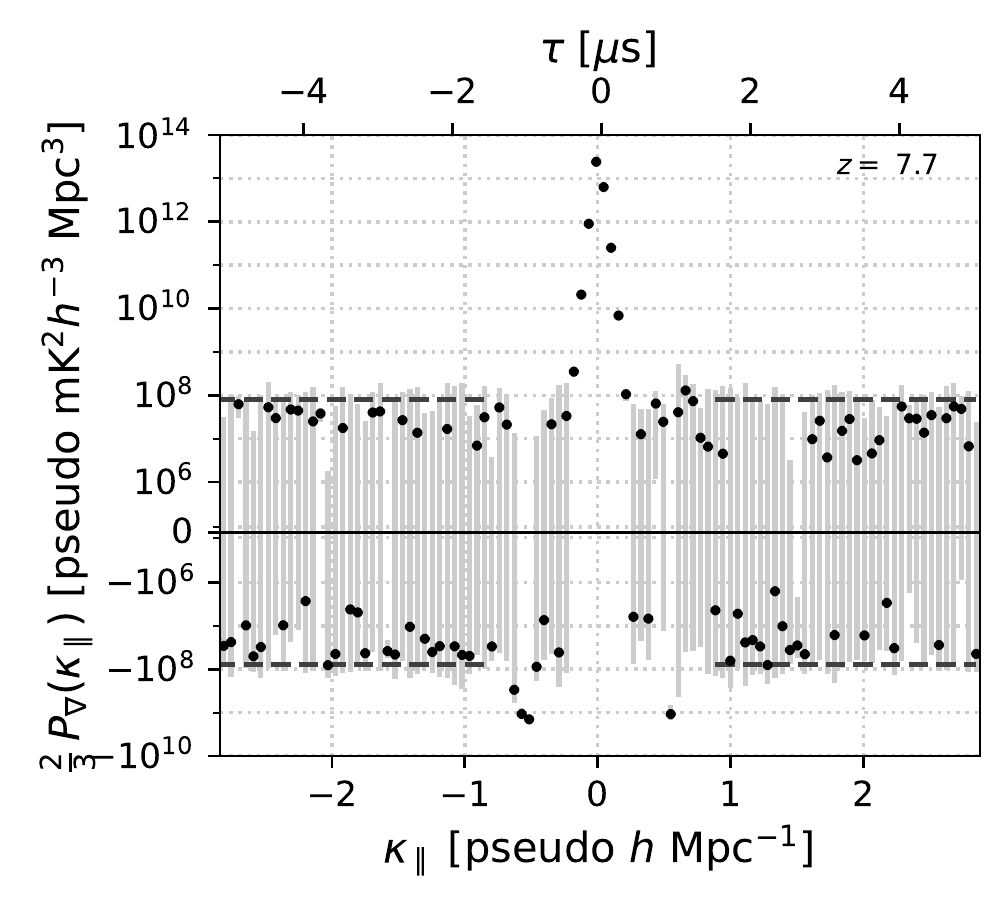}}
% \subfloat[][Data\label{fig:EQ28XX_data_RA1.6hr_z_7.7_triadcomb_2}]{\includegraphics[width=0.47\textwidth]{figures/EQ28XX_data_RA_0136_hr_transit_cpdps_z_7_median_resampled_dlst_60s_triadcomb_2.pdf}}
\subfloat[][Model\label{fig:EQ28XX_model_RA1.6hr_z_7.7_triadcomb_2}]{\includegraphics[width=0.47\textwidth]{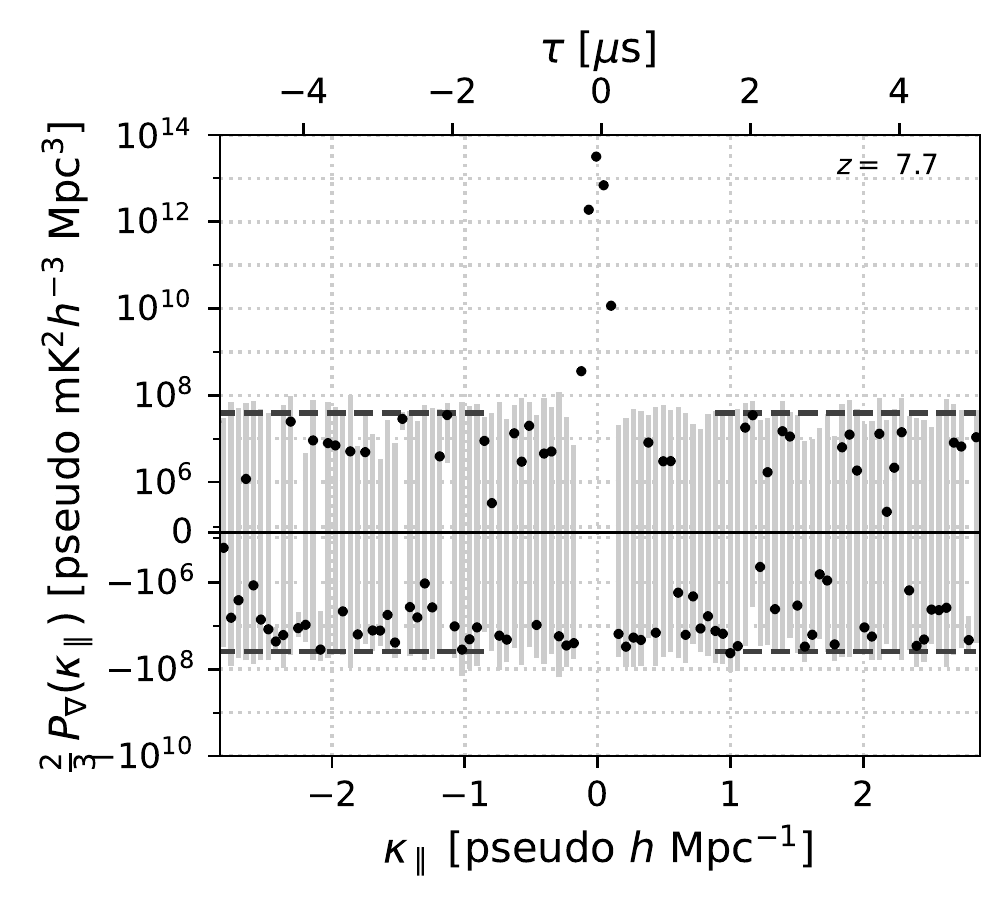}}
% \subfloat[][Model\label{fig:EQ28XX_model_RA1.6hr_z_7.7_triadcomb_2}]{\includegraphics[width=0.47\textwidth]{figures/EQ28XX_model_RA_0136_hr_transit_cpdps_z_7_median_resampled_dlst_60s_triadcomb_2.pdf}}
\caption{Same as Fig.~\ref{fig:EQ28XX-baseline-cpdps-RA1.6hr-z-7-median} but the values denote the cross-power from different pairs of triads falling under the same class (inter-triad). The usage of inter-triad cross-power results in a substantial mitigation of the secondary peaks at $|\kappa_\parallel|\simeq 0.5\,\textrm{``pseudo''}\,h$~Mpc$^{-1}$. This signifies that the systematic effect varies randomly between different triads and averaging of cross-power can be used to mitigate it substantially. The bib-to-bin variation at $|\kappa_\parallel|\gtrsim 0.85\,\textrm{``pseudo''}\,h$~Mpc$^{-1}$ has \textit{rms} $\simeq 7.9\times 10^7$~``pseudo''~mK$^2(\textrm{Mpc}/h)^3$ and $\simeq 5\times 10^7$~``pseudo''~mK$^2(\textrm{Mpc}/h)^3$ in the data and the model respectively. The two-sided K-S test indicates that the statistical discrepancy between the data and the model is only marginal and below the significance threshold. This indicates that the systematics present in intra-triad averaging are now eliminated to a significant extent by inter-triad averaging. The marginal discrepancies could be due to potentially underestimating the noise in the model. 
\label{fig:EQ28XX-upper-cpdps-RA1.6hr-z-7-median}}
\end{figure*}

Figure~\ref{fig:EQ28XX-upper-cpdps-RA1.6hr-z-7-median} shows the inter-triad cross-power spectrum for the J0136-30 field in the $z=7.7$ sub-band. In contrast with the findings noted in Fig.~\ref{fig:EQ28XX-baseline-cpdps-RA1.6hr-z-7-median}, the secondary peaks are suppressed significantly to $\lesssim 10^9$~``pseudo''~mK$^2(\textrm{Mpc}/h)^3$ at $|\kappa_\parallel|\simeq 0.5\,\textrm{``pseudo''}\,h$~Mpc$^{-1}$ indicating a noise-like limitation in the uncertainty ranges in virtually all except a few innermost $\kappa_\parallel$-bins, which is confirmed by the modeling (right panel). 

Besides the systematics, the overall uncertainties are also significantly reduced. For instance, the bin-to-bin variation at $|\kappa_\parallel|\gtrsim 0.85\,\textrm{``pseudo''}\,h$~Mpc$^{-1}$ has \textit{rms} $\simeq 7.9\times 10^7$~``pseudo''~mK$^2(\textrm{Mpc}/h)^3$ and $\simeq 5\times 10^7$~``pseudo''~mK$^2(\textrm{Mpc}/h)^3$ in the data and the model respectively. The discrepancy between the two is significantly lesser compared to the intra-triad case. This is only a marginal discrepancy as supported by the two-sample K-S metrics $D_\textrm{KS}\approx 0.135, P_\textrm{KS}\simeq 0.04\%$ which shows that the null hypothesis that the samples in the data and the model could be rejected only with marginal significance as $P_\textrm{KS}$ is less than the significance threshold. Potential contributions to this marginal discrepancy could be due to either mis-modeling of the noise (marginally underestimated noise level) or the usage of a few more triads in the model than in the data.

Again, the usage of more triads in the model compared to data implies the uncertainty estimates from the models are, in the worst case, likely to be as low as $\approx 84$\% ($\approx 31/37$) of the actual thermal uncertainties when the same number of triads are used. This is different than the intra-triad case because the inter-triad cross-power spectrum assumes an implicit redundancy and coherence between the triad pairs off the main diagonal ($\delta\nabla > 0$), whereas the intra-triad cross-power spectrum assumes no coherence and only uses the main diagonal, $\delta\nabla=0$. This correction to the estimated uncertainties is still insignificant considering the much higher dynamic range required to detect the fiducial \HI\ signal from the EoR.

\subsection{Sensitivity Improvements}\label{sec:Sensitivity-improvement}

If the power spectrum errors are incoherent, the sensitivity can be further improved by incoherent averaging of cross-power spectra obtained with different polarizations and locations on the sky, and binning adjacent $\kappa_\parallel$-modes into larger bins. In this paper, we have performed a simplified version of such an averaging over two polarizations (`XX' and `YY') and folding the negative- and positive-valued  $\kappa_\parallel$-modes into bins of $|\kappa_\parallel|$. The latter is motivated by assuming statistical isotropy of the cosmological signal, wherein the cosmological signal power in a $\kappa_\parallel$-mode should only depend on $|\kappa_\parallel|$.

\begin{figure*}
\centering
  \subfloat[][Dual-pol average of data\label{fig:EQ28_dualpol_data_cpdps_RA1.6hr_z_7.7_triadcomb_2}]{\includegraphics[width=0.45\linewidth]{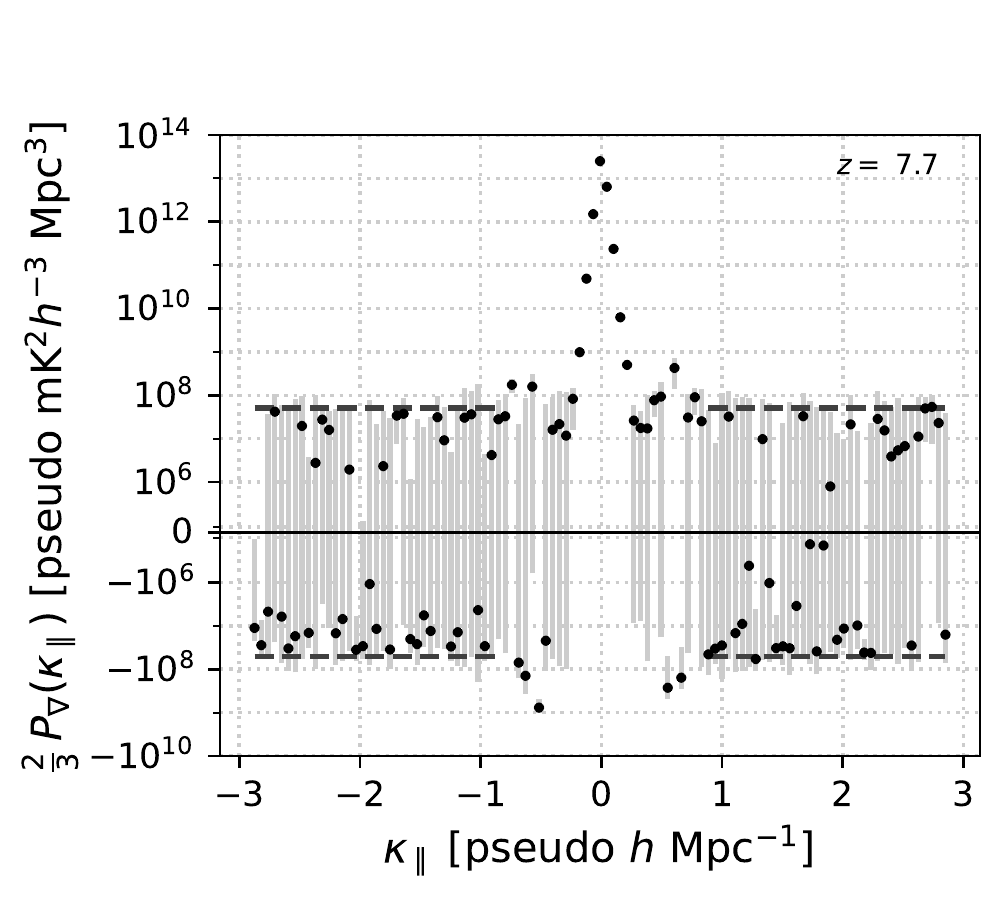}}
  % \subfloat[][Dual-pol average of data\label{fig:EQ28_dualpol_data_cpdps_RA1.6hr_z_7.7_triadcomb_2}]{\includegraphics[width=0.45\linewidth]{figures/EQ28_dualpol_data_avg_cpdps_z_7_median_resampled_dlst_60s_triadcomb_2.pdf}}
  \hspace{0.2cm}
  \subfloat[][Dual-pol and $|\kappa_\parallel|$ average of data \label{fig:EQ28_dualpol_data_folded_cpdps_RA1.6hr_z_7.7_triadcomb_2}]{\includegraphics[width=0.45\linewidth]{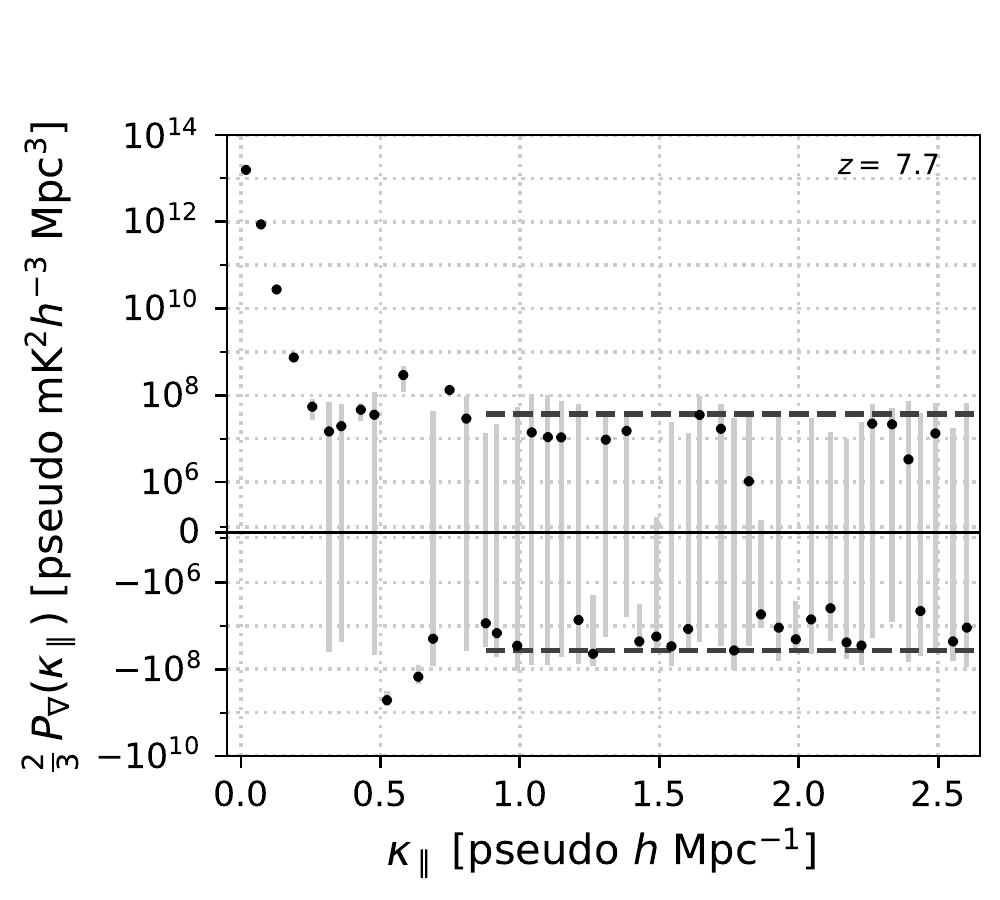}}
  % \subfloat[][Dual-pol and $|\kappa_\parallel|$ average of data \label{fig:EQ28_dualpol_data_folded_cpdps_RA1.6hr_z_7.7_triadcomb_2}]{\includegraphics[width=0.45\linewidth]{figures/EQ28_dualpol_data_kbin_avg_cpdps_z_7_median_resampled_dlst_60s_triadcomb_2.pdf}}
\caption{\textit{Left:} Same as Fig.~\ref{fig:EQ28XX_data_RA1.6hr_z_7.7_triadcomb_2} but the power spectra from `XX' and `YY' polarizations from the data have been incoherently averaged. The bin-to-bin \textit{rms} at $|\kappa_\parallel| \gtrsim 0.85\,\textrm{``pseudo''}\,h$~Mpc$^{-1}$ is $\simeq 5.1\times 10^7$~``pseudo''~mK$^2(\textrm{Mpc}/h)^3$, which is lower than that in the `XX' and `YY' polarizations by factors of $1.55$ and $1.25$ respectively. \textit{Right:} The dual-pol average of the data (left) is folded and averaged in bins of $|\kappa_\parallel|$. The bin-to-bin \textit{rms} of $\simeq 3.7\times 10^7$~``pseudo''~mK$^2(\textrm{Mpc}/h)^3$ at $|\kappa_\parallel| \gtrsim 0.85\,\textrm{``pseudo''}\,h$~Mpc$^{-1}$ signifies an improvement in sensitivity by a factor of $1.38$ relative to that from only averaging the two individual polarized cross-power spectra (left). A reduction in the secondary peaks is also noted relative to those in Fig.~\ref{fig:EQ28XX_data_RA1.6hr_z_7.7_triadcomb_2} signifying the incoherent nature of the systematic errors. \label{fig:EQ28XX-folded-upper-cpdps-RA1.6hr-z-7-median}}
\end{figure*}

Figure~\ref{fig:EQ28_dualpol_data_cpdps_RA1.6hr_z_7.7_triadcomb_2} shows the inter-triad cross-power spectrum of data in the J0136-30 field in the $z=7.7$ sub-band after averaging the data `XX' and `YY' polarizations. The bin-to-bin \textit{rms} at $|\kappa_\parallel|\gtrsim 0.85\,\textrm{``pseudo''}\,h$~Mpc$^{-1}$ of the `XX' and `YY' polarizations are $\simeq 7.9\times 10^7$~``pseudo''~mK$^2(\textrm{Mpc}/h)^3$ (see Fig.~\ref{fig:EQ28XX_data_RA1.6hr_z_7.7_triadcomb_2}) and $\simeq 6.4\times 10^7$~``pseudo''~mK$^2(\textrm{Mpc}/h)^3$ (not shown here) respectively. The corresponding uncertainty from averaging the power in the two polarizations is $\simeq 5.1\times 10^7$~``pseudo''~mK$^2(\textrm{Mpc}/h)^3$. This incoherent averaging results in an improvement in sensitivity by a factor of $\approx 1.55$ and $\approx 1.25$ relative to that in the `XX' and `YY' polarizations respectively, in line with that expected from averaging two independent realizations of noise.

Figure~\ref{fig:EQ28_dualpol_data_folded_cpdps_RA1.6hr_z_7.7_triadcomb_2} shows the inter-triad cross-power spectrum averaged between the two polarizations followed by folding along $\kappa_\parallel = 0\,\textrm{``pseudo''}\,h$~Mpc$^{-1}$ and further averaging in bins of $|\kappa_\parallel|$. The \textit{rms} at $|\kappa_\parallel|\gtrsim 0.85\,\textrm{``pseudo''}\,h$~Mpc$^{-1}$ reduces further to $\simeq 3.7\times 10^7$~``pseudo''~mK$^2(\textrm{Mpc}/h)^3$ (dashed lines) which signifies an improvement by a factor $\approx 1.38$ in sensitivity that could be expected from twice the number of samples because the noise is uncorrelated in the mirrored $|\kappa_\parallel|$-bins. The secondary peaks are also reduced in comparison to those in Fig.~\ref{fig:EQ28XX_data_RA1.6hr_z_7.7_triadcomb_2}. This demonstrates that expected improvements in sensitivity can be obtained by further averaging nominally redundant power spectra but obtained with independent realizations. The positive and negative fluctuations of the noise realizations but with a uniform noise floor at $|\kappa_\parallel| \gtrsim 0.85\,\textrm{``pseudo''}\,h$~Mpc$^{-1}$ shows that the analysis presented here is predominantly limited by thermal noise in the measurements rather than systematics, and that further improvements in sensitivity are possible with more data. Even in cases where the presence of systematics were noted to be comparable or in excess of the thermal noise contribution, the uncertainties still appear to be data-limited and not systematic-limited wherein additional data could further improve the sensitivity.

In most of this analysis, especially as seen in inter-triad cross-power spectra, we find that the uncertainty ranges in the data are attributable to thermal-like noise as supported by the corresponding models. The modeling fully includes the imperfect placement of antennas relative to an ideal grid. The averaging steps implicitly assume that only identical $k$-modes are processed, whereas in practice, information across non-identical modes are combined together as a result of implicit non-redundancy in the antenna spacings. This may add systematic uncertainties to the noise-like uncertainties. We have performed a test wherein the models contained an intentional underestimate of the noise \textit{rms} via the $T_\textrm{sys}$ model and we still found the uncertainty ranges to be correspondingly smaller and proportional to the square of the noise in the visibilities, thereby strongly implying that the antenna position imperfections do not contribute significantly to the uncertainties in the power spectra.

\begin{figure}
    \centering
    \includegraphics[width=0.95\linewidth]{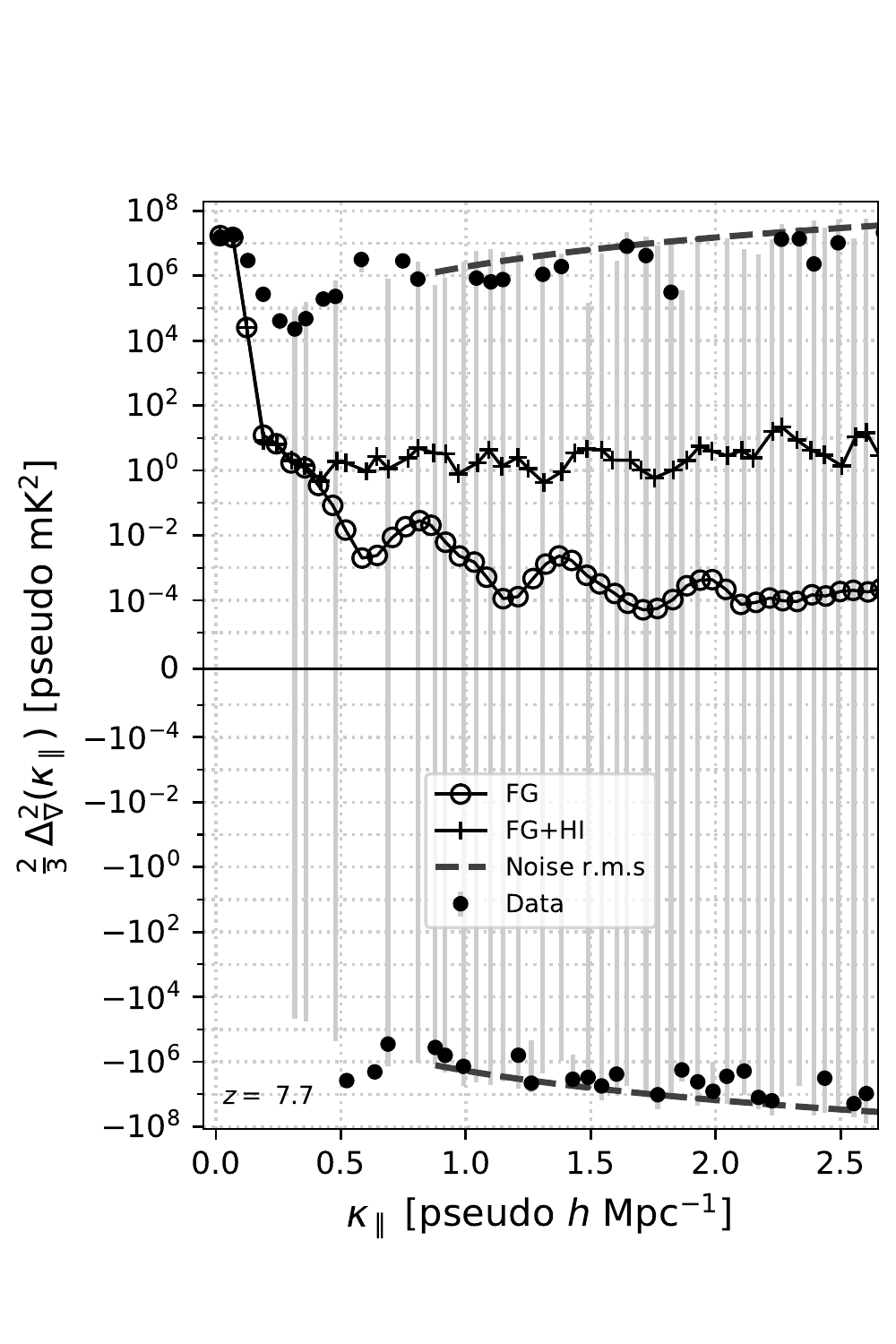}
    \caption{Same as Fig.~\ref{fig:EQ28_dualpol_data_folded_cpdps_RA1.6hr_z_7.7_triadcomb_2} but using $\Delta_\nabla^2(\kappa_\parallel)$, in units of ``pseudo''~mK$^2$, instead of $P_\nabla(\kappa_\parallel)$. The filled circles denote the data with error bars shown using vertical gray lines. The dashed lines denote the bin-to-bin \textit{rms} of noise power. The open circles denote the power spectrum from a foreground-only model of the J0136-30 field. The `plus' symbols denote the power due to the presence of EoR \HI\ fluctuations in excess over the foreground power. The best data- and noise-limited ``$1\sigma$'' upper limits are found to be $\le 316$~``pseudo''~mK and $\le 1000$~``pseudo''~mK at $\kappa_\parallel=0.33\,\textrm{``pseudo''}\,h$~Mpc$^{-1}$ and $\kappa_\parallel=0.875\,\textrm{``pseudo''}\,h$~Mpc$^{-1}$ respectively. \label{fig:EQ28_dualpol_data_folded_cpDel2_RA1.6hr_z_7.7_triadcomb_2}}
\end{figure}

Borrowing from cosmology convention, we can define the mathematical analog of the standard cosmological variance, $\Delta_\nabla^2(\kappa_\parallel) = \kappa^3\,P_\nabla(\kappa_\parallel)/(2\pi^2)$, with $\kappa^2=\kappa_\perp^2+\kappa_\parallel^2$ and $\kappa_\perp=2\pi (|\boldsymbol{b}_p|/\lambda)/D$ corresponding to $|\boldsymbol{b}_p|=29.2$~m. Figure~\ref{fig:EQ28_dualpol_data_folded_cpDel2_RA1.6hr_z_7.7_triadcomb_2} is equivalent to Fig.~\ref{fig:EQ28_dualpol_data_folded_cpdps_RA1.6hr_z_7.7_triadcomb_2} but uses $\Delta_\nabla^2(\kappa_\parallel)$, in units of ``pseudo''~mK$^2$, instead of $P_\nabla(\kappa_\parallel)$. Also shown is the power spectrum from a model containing only foregrounds in the J0136-30 field (open circles), and the modeled power due to EoR \HI\ fluctuations (`plus' symbols) in excess over that from foregrounds. By re-converting $\Delta_\nabla^2(\kappa_\parallel)$ in Fig.~\ref{fig:EQ28_dualpol_data_folded_cpDel2_RA1.6hr_z_7.7_triadcomb_2} to $P_\nabla(\kappa_\parallel)$, it can be shown by comparison to a regular visibility delay power spectrum (see Fig.~\ref{fig:delay-PS-29m-J0136-30-z7}) that the dynamic range required to separate the \HI\ signal from the peak foreground power are very similar.

The lowest-valued data-limited ``$1\sigma$'' upper limit is seen to be $(2/3)\Delta_\nabla^2(\kappa_\parallel) \le 10^5$~``pseudo''~mK$^2$ at $\kappa_\parallel=0.33\,\textrm{``pseudo''}\,h$~Mpc$^{-1}$. This corresponds to $\le 316$~``pseudo''~mK in temperature units. However, the neighboring data points are not data-limited and exhibit a `detection' well above noise. More conservatively, in the predominantly noise-like regions beyond $|\kappa_\parallel| \gtrsim 0.85\,\textrm{``pseudo''}\,h$~Mpc$^{-1}$, a measurement with the least amount of uncertainty range can be seen at $\kappa_\parallel=0.875\,\textrm{``pseudo''}\,h$~Mpc$^{-1}$ with a ``$1\sigma$'' upper limit of $(2/3)\Delta_\nabla^2\le 10^6$~``pseudo''~mK$^2$, or equivalently, $\le 1000$~``pseudo''~mK in temperature units. 

We re-emphasize that the physical validity of either directly attributing this temperature to the brightness temperature power spectrum of the IGM \HI\ 21cm signal or interpreting $\kappa_\parallel$ to a distance scale is only approximate. The interpretation of $\Delta_\nabla^2(\kappa_\parallel)$ will not be a straightforward analog of the standard $\Delta^2(k_\parallel)$ because the nature of the fluctuations in bispectrum phase is not identical to that in the standard visibilities. However, the two are approximately related and this relationship is mathematically established in Paper~I. 

Our initial goal is a simple detection of the EoR \HI\ fluctuations in the form of excess variance in bispectrum phase over that expected from just smooth-spectrum radio foregrounds. The analysis presented here establishes a means for detecting cosmic \HI\ while distinguishing it from the foregrounds through a robust quantity, the bispectrum phase, and have used techniques that parallel other standard approaches such as the \textit{delay spectrum}. 

\section{Summary}\label{sec:Summary}

While numerous improvements in various calibration schemes are being devised, we provide here the first demonstrative analysis using a novel technique using bispectrum phase, presented in \cite{thy18}, that serves as an independent alternative to existing approaches that are variants of a standard power spectrum. The bispectrum phase angle is a measure of the `symmetry' of the spatial distribution of the brightness. The fluctuations in bispectrum phase angle as sought in this paper measure the dissimilarity of the EoR \HI\ signal relative to the foregrounds. The key benefits of the approach presented here are that the bispectrum phase is a robust sky measurement, immune to antenna-based direction-independent calibration and associated errors, and that established techniques that are used for spectral discrimination of the EoR signal such as the \textit{delay spectrum} can be readily employed. We have illustrated that the spectral dynamic range required to detect the fluctuations from the redshifted \HI\ signal relative to the foregrounds is very similar to that in other standard approaches.  

In this paper, we have used a small subset of data obtained using the HERA telescope (50 out of 61 antennas in this analysis), $\approx 22$~min in duration centered on each of the two fields (Fornax~A transit and J0136-30) and repeatedly observed over 18 nights. We chose a relatively RFI-free frequency sub-band with the band center corresponding to $z=7.7$ and equilateral triads made of 29.2~m antenna spacings. Interpreting results using bispectrum phase requires detailed forward-modeling, a limited example of which we have presented here to support the data analysis including a fiducial model for the EoR from 21cmFAST simulations, and the most accurate foreground continuum sky models currently available. Using a parallel analysis with detailed modeling, we are able to confirm that the analysis on the amount of data presented here is currently noise-limited in some cases studied here and definitely data-limited in most cases. Being a robust diagnostic of baseline-dependent errors, the bispectrum phase approach distinctly identifies potential baseline-dependent systematics at $|\kappa_\parallel|\simeq 0.5\,\textrm{``pseudo''}\,h$~Mpc$^{-1}$ in the HERA data used in this analysis. It must be noted that the improvement further required in dynamic range or sensitivity is very similar to that required from other standard analysis techniques that fall under the \textit{foreground avoidance} category, while re-emphasizing that the results presented here neither required antenna calibration nor sophisticated analysis methods. 

Drawing from cosmological precedent, the existence of parallels with the \textit{delay spectrum} approach allow us to define the variance of `temperature' fluctuations, $\Delta_\nabla^2(\kappa_\parallel)$. The best data-limited ``$1\sigma$'' upper limit in temperature units was found to be $\le 316$~``pseudo''~mK at $\kappa_\parallel=0.33\,\textrm{``pseudo''}\,h$~Mpc$^{-1}$. More conservatively, in noise-limited modes, $\kappa_\parallel\gtrsim 0.85\,\textrm{``pseudo''}\,h$~Mpc$^{-1}$, the best ``$1\sigma$'' upper limit was found to be $\le 1000$~``pseudo''~mK at $\kappa_\parallel=0.875\,\textrm{``pseudo''}\,h$~Mpc$^{-1}$. We note that these results are to be interpreted only in an approximate sense as physical brightness temperatures and distance scales, because the origin of fluctuations in the bispectrum phase, while approximately related, are not the same as that in the standard two-point spatial coherence.

Our data-limited results indicate that there is scope for improving sensitivity further with more data. When fully built, HERA will have 350 dishes and will be able to observe round the year a $\simeq 10$\arcdeg-wide stripe at $\delta=-30\fdg7$. This will yield $\sim 10$ times more nominally redundant antenna spacings, and $\sim 16$ independent patches on the sky over 12~hours of observing per night. Ideally, assuming no new significant systematic effects are encountered, we anticipate a full-sized data set to yield sufficient sensitivity to be able to place useful constraints on fiducial EoR models even if a detection remains elusive.

Our bispectrum phase approach is designed as an independent alternative to detect and confirm the presence of excess fluctuations due to the presence of \HI\ during the EoR relative to a null hypothesis when such signals are absent, rather than provide a direct interpretation of the astrophysical conditions. Astrophysical interpretation will require extensive forward-modeling. Such a forward-modeling will allow us to relate the bispectrum phase fluctuations to the magnitude of the brightness temperature fluctuations and the astrophysical conditions of the IGM.

\begin{acknowledgments}
This material is based upon work supported by the National Science Foundation under Grants No. 1636646 and No. 1836019 and institutional support from the HERA collaboration partners.  This research is funded in part by the Gordon and Betty Moore Foundation. HERA is hosted by the South African Radio Astronomy Observatory, which is a facility of the National Research Foundation, an agency of the Department of Science and Innovation. The National Radio Astronomy Observatory is a facility of the National Science Foundation operated under cooperative agreement by Associated Universities, Inc. AL acknowledges support from the New Frontiers in Research Fund Exploration grant program, a Natural Sciences and Engineering Research Council of Canada (NSERC) Discovery Grant and a Discovery Launch Supplement, the Sloan Research Fellowship, the William Dawson Scholarship at McGill, as well as the Canadian Institute for Advanced Research (CIFAR) Azrieli Global Scholars program. We acknowledge softwares including Python, Numpy, SciPy, Astropy, Matplotlib, MPI for Python, HEALPix, Healpy that made the numerical computations and the creation of figures presented in this manuscript possible. 
\end{acknowledgments}

\appendix

\section{Thermal Noise Uncertainty in Standard Visibility Delay Power Spectrum}\label{sec:analytical-thermal-noise-power}

Here, we quasi-analytically derive the \textit{standard deviation} of thermal noise uncertainty in the standard delay spectrum of visibilities under different assumptions of redundancy in measurements. Consider different types of pairs of antennas of given spacing but different orientations. If there are $N_{bj}$ antenna pairs of each type $j$ all with the same magnitude of spacing, the total number of antenna pairs with this spacing is $N_b = \sum_j N_{bj}$. 

The data are assumed to have been measured repeatedly over $N_r^\prime$ nights covering $N_\textrm{fld}$ fields for a duration, $t_\textrm{fld}$, each night exactly centered on the transit of each field covering the same range of LST. From each of these nights, the data are assumed to be coherently averaged for a time interval of $t_\textrm{int}$, thus leaving $N_l=t_\textrm{fld}/t_\textrm{int}$ temporal bins on each field each of temporal width $t_\textrm{int}$. To avoid potential noise bias in the delay power spectrum, instead of averaging all the LST-aligned nights together, the data are split into $N_r$ chunks comprising of equal number of nights in each, $N_n=N_r^\prime/N_r$, and averaged between these split nights while assuming perfect LST alignment across them. Thus, these equally split chunks are expected to have independent but identical Gaussian distributions for the noise. 

The delay power spectrum is obtained from averaging the real parts of the cross-multiplications of the complex-valued delay spectra from different pairs of chunks, $N_{rr}=N_r\,(N_r-1)/2$, after accounting for the cosmological scaling factors that translate power in an observer's coordinates to cosmological coordinates \cite{mor05,mcq06,par12a}. If redundancy is assumed, the visibilities measured on antenna spacings in different orientations can be averaged coherently within each type. Otherwise, the delay power spectra on each antenna spacing can be individually computed and averaged together in an incoherent sense assuming statistical isotropy of the cosmological \HI\ signal. The former will yield a higher sensitivity on account of coherent averaging within the same type of antenna spacing, but may be more severely subject to systematics than the latter. 

The uncertainty due to thermal noise in the delay power spectrum can be obtained analytically as follows. The thermal noise \textit{rms} on a measured visibility amplitude after averaging over all nights in a chunk is $\sigma_\textrm{n}=\sigma_\textrm{T}/\sqrt{\Delta f\,t_\textrm{int}\,N_n}$. It can be shown that the thermal noise uncertainty in each bin of the delay spectrum of the visibilities has an \textit{rms} of magnitude $\sigma_\textrm{d}=\sigma_\textrm{n}\sqrt{\Delta B/\Delta f}\,\Delta f$ \cite{mor05,par12a}. The real and imaginary parts of noise in each bin of the delay spectrum are independent and identical Gaussian random variables with an \textit{rms} of $\sigma_\textrm{d}/\sqrt{2}$. 

When two independent Gaussian random variables, $X$ and $Y$, are multiplied, the resulting distribution is a modified Bessel function of the second kind \cite{cra36,nad16}. In our case, the the resulting cross-product is a complex variable whose real and imaginary parts each follow this distribution. It can be numerically demonstrated that the real and imaginary parts of the cross-product have respective variances of $\sigma_\textrm{RR}^2=\sigma_\textrm{R1}^2\sigma_\textrm{R2}^2 + \sigma_\textrm{I1}^2\sigma_\textrm{I2}^2$, and $\sigma_\textrm{II}^2=\sigma_\textrm{R1}^2\sigma_\textrm{I2}^2 + \sigma_\textrm{I1}^2\sigma_\textrm{R2}^2$, where $\textrm{R}$ and $\textrm{I}$ correspond to real and imaginary parts respectively, and the numerical index denotes the individual random variable in the cross-product. 

With the availability of $N_{rr}$ pairs of chunks, the thermal noise \textit{rms} in the power spectrum, $\sigma_\textrm{RR}$, can be potentially reduced via incoherent averaging. However, if $N_r>2$ or $N_{rr}>1$, the power spectrum realizations across these pairs of chunks may be correlated, thus not offering much gain in sensitivity. Hence, for the simple example in this paper, we only consider $N_r=2$ (or $N_{rr}=1$) hereafter. Averaging the power spectra from $N_\textrm{fld}$ fields and $N_l$ adjoining temporal bins of LST in each field will further improve sensitivity by a factor $\sqrt{N_\textrm{fld}\,N_l}$. If independent modes of polarizations (up to two orthogonal modes, $N_\textrm{pol}\in\{1,2\}$) are also available for incoherent averaging of the power spectra, the thermal noise \textit{rms} in the power spectrum can be reduced to $\sigma_\textrm{RR}/\sqrt{N_\textrm{fld}\,N_l\,N_\textrm{pol}}$. Finally, the weights from all available baselines is considered under two scenarios -- redundant and non-redundant. It can be shown that the \textit{rms} in the power spectrum can be reduced to $\sigma_\textrm{RR}/\sqrt{N_\textrm{fld}\,N_l\,N_\textrm{pol}\,w_b}$, where, $w_b=\sum_j N_{bj}^2$ or $w_b=\sum_j N_{bj}$ for the redundant and non-redundant cases respectively. As the number of independent power spectrum realizations increase, such an average will approach a Gaussian distribution according to the \textit{Central Limit Theorem}. 

By making the simplifying assumption that $\sigma_\textrm{R1}=\sigma_\textrm{R2}=\sigma_\textrm{I1}=\sigma_\textrm{I2}=\sigma_\textrm{d}/\sqrt{2}$, the thermal noise \textit{rms} in the simple delay cross-power spectrum that is free of noise-bias can be written as:
\begin{align}
    P_\textrm{N} &= \frac{\sigma_\textrm{RR}}{\sqrt{N_\textrm{fld}\,N_l\,N_\textrm{pol}\,w_b}}\left(\frac{A_\textrm{e}}{\lambda^2\Delta B}\right)\left(\frac{D^2\Delta D}{\Delta B}\right)\left(\frac{\lambda^2}{2\,k_\textrm{B}}\right)^2 \nonumber\\
    &= \frac{\lambda^2}{A_\textrm{e}}\cdot \frac{D^2\Delta D}{\sqrt{2\, N_\textrm{fld} \, N_\textrm{pol}\,w_b}}\cdot \frac{T_\textrm{sys}^2}{\Delta B\,(N_r^\prime/2)\,\sqrt{t_\textrm{int}t_\textrm{fld}}},
\end{align}
where, $\lambda$ is the wavelength of the band center, $D\equiv D(z)$ is the comoving distance to redshift $z$, $\Delta D$ is the comoving depth along the line of sight corresponding to $\Delta B$, and $k_\textrm{B}$ is the Boltzmann constant.

For the observing parameters and the 29.2~m antenna spacings considered here, the HERA layout described above yields $N_b=124$ pairs of antennas spaced 29.2~m apart with $N_{b1}=41$, $N_{b2}=40$, and $N_{b3}=43$ antenna pairs oriented at 0\arcdeg, 60\arcdeg, and $-60$\arcdeg~ relative to due east respectively. In our example, $N_r^\prime = 18$ and we choose $N_r=2$ with $N_n=9$. On each night, $t_\textrm{fld}=22$~min with $N_\textrm{fld}=1$, and we choose a coherent averaging interval of $t_\textrm{int}=1$~min. $P_\textrm{N}$ obtained with and without assumption of redundancy of antenna spacings is shown in Fig.~\ref{fig:delay-PS-29m-J0136-30-z7}.

% \bibliography{eor}

\end{document}